\documentclass[twocolumn,trackchanges,twocolappendix]{aastex63}

\usepackage{natbib,enumitem,array,amsmath}

\graphicspath{{./}{figures/}}

\newcommand{\btwnpoint}{\hspace{-0.55ex}}

\newcommand{\pointsec}{%
  \mathrel{\vbox{\offinterlineskip\ialign{%
    ##\cr
    $\btwnpoint\scriptstyle\text{s}$\cr
    \noalign{\kern0.75ex}
    $\btwnpoint.$\cr
}}}}

\newcommand{\pointarcsec}{%
  \mathrel{\vbox{\offinterlineskip\ialign{%
    ##\cr
    $\btwnpoint\scriptstyle\prime\prime$\cr
    \noalign{\kern0.75ex}
    $\btwnpoint.$\cr
}}}}

\definecolor{edit_color}{HTML}{000000} 
\newcommand{\myedit}[1]{\textcolor{edit_color}{#1}}

\received{22 December 2020}
\revised{10 March 2021}
\accepted{26 March 2021}
\submitjournal{AJ}

\shorttitle{Substructure Quantification in Dwarf ETGs} 
\shortauthors{Michea et al.}

\begin{document}

\def\arraystretch{1.13} 
\setlength{\tabcolsep}{0.5em}

\title{Brought to Light I: Quantification of Disk Substructure in Dwarf Early-Type Galaxies}

\correspondingauthor{Josefina Michea}
\email{michea@ari.uni-heidelberg.de}

\author[0000-0001-8205-945X]{Josefina Michea}
\affiliation{Astronomisches Rechen-Institut, Zentrum f\"ur Astronomie der Universit\"at Heidelberg, M\"onchhofstra{\ss}e 12-14, 69120 Heidelberg, Germany}

\author[0000-0001-5171-5629]{Anna Pasquali}
\affiliation{Astronomisches Rechen-Institut, Zentrum f\"ur Astronomie der Universit\"at Heidelberg, M\"onchhofstra{\ss}e 12-14, 69120 Heidelberg, Germany}

\author[0000-0001-5303-6830]{Rory Smith}
\affiliation{Korea Astronomy and Space Science Institute, 776 Daedeokdae-ro, Yuseong-gu, Daejeon 34055, Republic of Korea}

\author[0000-0001-6180-0245]{Katarina Kraljic}
\affiliation{Aix Marseille Universit\'e, CNRS, CNES, UMR 7326, Laboratoire d'Astrophysique de Marseille, Marseille, France}
\affiliation{Institute for Astronomy, University of Edinburgh, Royal Observatory, Blackford Hill, Edinburgh, EH9 3HJ, United Kingdom}

\author[0000-0002-1891-3794]{Eva K. Grebel}
\affiliation{Astronomisches Rechen-Institut, Zentrum f\"ur Astronomie der Universit\"at Heidelberg, M\"onchhofstra{\ss}e 12-14, 69120 Heidelberg, Germany}

\author[0000-0002-7069-113X]{Paula Calder\'on-Castillo}
\affiliation{Departamento de Astronom\'ia, Universidad de Concepci\'on, Casilla 160-C, Concepci\'on, Chile}

\author[0000-0002-6807-5856]{Thorsten Lisker}
\affiliation{Astronomisches Rechen-Institut, Zentrum f\"ur Astronomie der Universit\"at Heidelberg, M\"onchhofstra{\ss}e 12-14, 69120 Heidelberg, Germany}

\begin{abstract} 
Dwarf early-type galaxies (ETGs) display a rich diversity in their photometric, structural, and dynamical properties. In this work, we address their structural complexity by studying with deep imaging a sample of nine dwarf ETGs from the Virgo galaxy cluster, characterized by having faint disk features, such as bars and spiral arms, which lie mostly hidden within the bright diffuse light of the galaxies. We present a new, robust method that aims to identify and extract the disk substructure embedded in these dwarf ETGs. The method consists in an iterative procedure that gradually separates a galaxy image into two components; the bright, dominant diffuse component, and the much fainter, underlying disk component. By applying it to the dwarf ETG sample, we quantify their disk substructure and find that its relative contribution to the total galaxy light ranges between 2.2 to 6.4\% within two effective radii. We test the reliability of the method, and prove that it is accurate in recovering the disk substructure we introduce in mock galaxy images, even at low disk-to-total light fractions of a few percent. As a potential application of the method, we perform a Fourier analysis on the extracted disk substructures and measure the orientation, length, and strength of the bars, and the pitch angle and strength of the spiral arms. We also briefly discuss a scenario based on the numerical simulations presented in our companion paper, Brought to Light II: Smith et al. 2021, in which we investigate the origins of the substructure in such dwarf systems.
\end{abstract}


\section{Introduction}\label{sec:introduction}

Under the paradigm of hierarchical structure formation, small-sized, low-mass galaxies are the most abundant galaxy type in the universe. In particular, dwarf early-type galaxies (ETGs) are commonly found in association with other galaxies in groups and clusters \citep{Binggeli:1985,Ferguson:1989}. Dwarf ETGs encompass dwarf elliptical \citep[dE;][]{Ferguson:1994}, dwarf lenticular or S0 \citep[dS0;][]{Sandage:1984}, and dwarf spheroidal \citep[dSph;][]{Grebel:2003} galaxies. Due to being dwarf systems, they are characterized by having a faint luminosity and surface brightness \citep[with the bright end being at $M_{B}\approx -18$ mag;][]{Ferguson:1994}, a low stellar mass ($\log M_{*}/M_{\odot} \lesssim 9.0-9.5$), and a small intrinsic size \citep[$R_{e} \lesssim 1-2$ kpc;][]{Eigenthaler:2018, Venhola:2019}. The early-type classification results from their general lack of star formation and their low gas and dust content \citep{Grebel:2001}, which leads to their red global optical color \citep{vanZee:2004b,Lisker:2008}. Additionally, dwarf ETGs are also mostly featureless in appearance, with a light distribution that is predominantly smooth \citep{Binggeli:1991}, \myedit{although many bright dwarf ETGs have also been found to host nuclei} \citep{Ordenes:2018,Venhola:2019}. Like their normal-sized elliptical counterparts, they tend to cluster towards high-density regions, thus extending the morphology-density relation to the low-mass galaxy regime \citep{Dressler:1980,Lisker:2007}. However, dwarf ETGs can also be quite heterogeneous, for example by presenting various degrees of rotational support \citep{Toloba:2015,Janz:2017,Bidaran:2020} and a wide range of ages and metallicities \citep{Jerjen:2004,Paudel:2010,Toloba:2014}, possibly pointing towards multiple origin and evolutionary scenarios.

Dwarf galaxies are important tracers of environmental build-up history, as their shallow potential well makes them especially susceptible to the environment. Consequently, they can be more easily affected by tidal harassment both from close fly-by galaxy encounters and from the gravitational potential of their host group or cluster, which can lead to mass stripping and morphological transformations \citep{Moore:1996,Smith:2010,Smith:2015}. In particular, \myedit{gas-rich, late-type dwarf galaxies can also experience starvation or strangulation when falling into a high-density region, where the hot, massive halo of the host group or cluster halts the gas inflow into the galaxies, causing them to eventually stop forming stars once their cold gas reservoirs are depleted} \citep{Larson:1980,Boselli:2008}. Similarly, their remaining gas can be removed by the hot intra-group or intra-cluster medium through ram-pressure stripping, effectively quenching their star formation \citep{Gunn:1972,vanZee:2004a,DeRijcke:2010}. Together, all these external effects can transform the properties of dwarf galaxies \myedit{from late to early types}, thus giving observable hints about the environmental context they have evolved in.

A highly suitable place to study dwarf galaxies in relation to their environment is the Virgo galaxy cluster, the nearest \myedit{\citep[16.5 Mpc;][]{Blakeslee:2009}} large-scale, \myedit{massive}, high-density agglomeration of galaxies \citep{Schindler:1999}. From the high- to the low-mass galaxy end, its members have been thoroughly catalogued and extensively studied; see, e.g., the ample scientific output of Virgo cluster surveys such as the Virgo Cluster Catalog \citep[VCC;][]{Binggeli:1985}, the Advanced Camera for Surveys Virgo Cluster Survey \citep[ACSVCS;][]{Cote:2004}, and the Next Generation Virgo Cluster Survey \citep[NGVS;][]{Ferrarese:2012}. \myedit{Interestingly, for a cluster of its mass \citep[$1.4-4.0\times10^{14}\, \text{M}_{\odot}$;][]{Weinmann:2011}, the Virgo cluster contains a consistently low fraction of ETGs across all galaxy masses \citep{Janz:2021}. This may be related to the fact that it is also a dynamically young cluster, showing} signatures of ongoing assembly such as field galaxies and galaxy groups that are currently infalling \citep{Binggeli:1987,Lisker:2018}. Paired with the gradual, hierarchical assembly of large structures, the Virgo cluster displays a striking contrast of galaxy populations, in which more evolved, \myedit{more gas-deficient} galaxies lie near the dense cluster core as they fell in earlier, while less evolved galaxies tend to be in the cluster outskirts as they are still infalling. This heterogeneity is particularly imprinted in the observable properties of the more vulnerable dwarf systems belonging to the cluster \citep[see, e.g., studies by][]{Lisker:2007,Paudel:2010,Toloba:2011}.

While it is true that a large fraction of dwarf ETGs can be characterized as having featureless, smooth light profiles that can be described by a single component, there are also cases that are structurally more complex, being best described by multiple components \citep{Janz:2012}. In such cases, substructure elements are often present: the central region of such galaxies can host nuclei and blue cores \citep{Lisker:2006b,Urich:2017,Hamraz:2019}, while their main body may show embedded weak disk features such as bars, spiral arms, rings, and dumbbells \citep{DeRijcke:2003,Graham:2003,Venhola:2018}. The majority of the dwarf ETGs with disk substructure that have been found and studied belong to the Virgo cluster, \myedit{owing to its close proximity} \citep[e.g.,][]{Jerjen:2000,Jerjen:2001,Barazza:2002,Geha:2003,Ferrarese:2006,Lisker:2006a,Lisker:2007,Lisker:2008,Janz:2012,Janz:2016}. Despite being widely studied, the origin of these disk features is still debated. One possibility is that they are remnant features of late-type progenitors that are transitioning to dwarf early-types through an environmental transformation, which has quenched their star formation and is in the process of changing the global morphology of the galaxies from disky to ellipsoidal \citep{Moore:1998,Penny:2014}. An opposite possibility is that the environment is not causing these disk features to fade away, but is instead triggering them in dwarf early-types \myedit{which could have been originally assembled, for example, through minor mergers. The formation of disk features could then be linked to the tidal harassment caused by other galaxies and the group or cluster potential \citep{Aguerri:2009,Gajda:2017,Kwak:2017,Kwak:2019}}. It is also possible that a high-density environment is not playing such a fundamental role, as dwarf ETGs with substructure have also been found in the field, and are believed to have formed through accretion events \citep{Graham:2017,Janz:2017}. These aforementioned scenarios are not mutually exclusive nor are they exhaustive, so the disk substructure that is observed in dwarf ETGs could potentially have multiple formation channels \citep[see, e.g., discussions by][]{deRijcke:2005,Lisker:2009b,Lisker:2012}.

Compared to the bright and diffuse main body of dwarf ETGs, any embedded disk features are much fainter, and thus lie mostly hidden from visual inspection. Therefore, image analysis techniques are necessary in order to detect and analyze these features. The simplest and most efficient way of revealing underlying substructure is through unsharp masking, in which a galaxy image is smoothed through convolution with a kernel function, and then the galaxy image is either divided by its smoothed-out version, or the smoothed-out version is subtracted from the galaxy image. Both approaches result in the removal of the dominant diffuse light, leaving behind the remaining substructure in an unsharp mask image \citep{McGaugh:1990,Erwin:2004,Lisker:2006c}. \myedit{More elaborate techniques involving unsharp masking have also been developed; for example, for the purpose of quantifying clumps typically associated with galaxy star-forming regions \citep{Conselice:2003}}. However, in general, unsharp masking can only be recommended for detection purposes, as the smoothing process inevitably redistributes the light of the galaxy, and thus alters both the actual light content and the resolution of the substructures.

\myedit{Another possibility is to adopt a modeling approach, and attempt to model the axisymmetric component of the galaxy. By subtracting the axisymmetric model from the galaxy image, the non-axisymmetric component containing the disk features, such as bars and spiral arms, would remain.} Several attempts at the characterization and quantification of the disk substructure in dwarf ETGs have followed this approach \citep[e.g.,][]{Barazza:2002,Lisker:2006a}. However, the modeling of the diffuse component tends to be simplistic, for example, by assuming a single shape and orientation \myedit{throughout all galactocentric radii}, which directly affects the accuracy with which the true underlying disk component is represented.

With the objective of improving on previous approaches, in this work we present a robust, newly developed method that aims to accurately identify and extract the disk substructure that is embedded in dwarf ETGs. The method is applied to a data sample that consists of deep imaging of nine dwarf ETGs from the Virgo cluster, which we introduce in Section \ref{sec:data}. In Section \ref{sec:method}, we describe the method in detail, explaining its steps and configuration. In Section \ref{sec:results}, we present the main results obtained by applying the method to the dwarf ETG sample, which allows us to accurately quantify and analyze their underlying disk features. To assess the reliability of the method, we construct a mock galaxy sample and subject it to diagnostic tests in Section \ref{sec:rectests}. Then, to show a practical application of the method, we carry out a Fourier analysis of the disk substructure of the dwarf ETG sample in Section \ref{sec:fourier}. In Section \ref{sec:discussion}, we discuss the strengths of the method and its potential applications. We also make a link to our companion paper, Brought to Light II: Smith et al. 2021, and address what we can learn through simulations of cluster harassment applied to dwarf galaxies. Finally, we provide a summary of this work in Section \ref{sec:summary}.

\section{Data}\label{sec:data}

The data sample analyzed in this work consists of nine Virgo cluster dwarf ETGs. These galaxies are certain cluster members according to the Virgo Cluster Catalog \citep[VCC;][]{Binggeli:1985}, and correspond to VCC0216, VCC0308, VCC0490, VCC0523, VCC0856, VCC0940, VCC1010, VCC1695, and VCC1896. \cite{Lisker:2006a} detected unambiguous disk features in these galaxies, based on unsharp masking of images from the Sloan Digital Sky Survey (SDSS) Data Release 4 \citep[DR4;][]{Adelman-McCarthy:2006}. In order to obtain deeper optical imaging, eight of them were targeted with the MPG/ESO 2.2m telescope at the European Southern Observatory (ESO), La Silla, using the Wide Field Imager \citep[WFI;][]{WFI:1999} instrument (observation program 077.B-0785, PI T. Lisker). The field of view centered on the dwarf galaxy VCC1010 also contained VCC0940, which thanks to apparent disk features was subsequently added to the sample.

With the purpose of maximizing the signal, the observations were taken in the white filter. The net exposure time aimed for each target was 2.5 hours, although for some targets this was not achieved due to bad weather. The observations were split into individual dithered exposures of 5 minutes, to minimize the saturation of bright stars and to allow the construction of a good superflat image during the data reduction. The data reduction was carried out with the THELI image data reduction pipeline \citep{THELI:2005,THELI:2013}, specifically designed for the MPG/ESO 2.2m WFI. We modified some of the procedures, in order to ensure that: \textit{(a)} the masks of extended sources were clearly covering all source pixels, and \textit{(b)} the image scaling for the superflat calculation correctly took into account these masks when computing mean or median image values. The latter required several iterations, constructing new object masks from a first iteration, and applying these in a second run.

The THELI pipeline also handled the registration of the individual images by using publicly available star catalogs, the shifting and rebinning by using a Lanczos kernel, and finally carrying out the relative flux calibration of the images before coadding. The final coadded and background-subtracted images reach a surface brightness that lies in the range of $27.2-27.9$ mag arcsec$^{-2}$ in the SDSS $r$-band equivalent, with a median depth across the sample of $\mu_{r} = 27.8$ mag arcsec$^{-2}$, which corresponds to $\mu_{B} = 29$ mag arcsec$^{-2}$ assuming typical dwarf ETG colors. The images were created using $3\times3$ pixel binning in order to reach the aforementioned depths at a signal-to-noise ratio (S/N) of 1 per binned pixel, with a scale equal to 0.71 arcsec per binned pixel. For the working sample, we constructed cutouts \myedit{of the $3\times3$ binned images} centered on each galaxy. \myedit{We note that this data reduction and final binned galaxy cutouts have also been used in the works of \citet{Lisker:2009c}, \citet{Lisker:2009a}, and \citet{Lisker:2009b,Lisker:2012}}.

Before beginning to work with the galaxy images, we carried out some additional analyses and pre-processing. For this purpose, we made use of the Image Reduction and Analysis Facility \citep[IRAF;][]{IRAF:1986, IRAF:1993}, an astronomical software system developed by the National Optical Astronomy Observatory (NOAO), and also of the SAOImage DS9 \citep{ds9:2003, ds9:2019} visualization tool for astronomical data. We used them to analyze each galaxy image separately. First, we measured the noise level of the background using the DS9 statistics tool by targeting regions that are devoid of any objects. We afterwards confirmed each of the measurements with the IRAF \texttt{imstat} task. The background noise of the sample is characterized by a standard deviation $\sigma_{\text{noise}}$ that ranges between $0.06-0.11$ counts s$^{-1}$ with a median value of $\sigma_{\text{noise}} = 0.07$ counts s$^{-1}$. Another important quantity to constrain is the point spread function (PSF) of the galaxy images. We visualized the images through DS9 to identify unsaturated foreground stars, and overlaid the IRAF \texttt{imexamine} task to measure their full width at half maximum (FWHM). Thus, we obtained that the PSF FWHM of the sample ranges between $1.13-1.41$ arcsec, with a median value of 1.27 arcsec. Both the background noise level and PSF FWHM of each galaxy image are relevant parameters that are used in later analyses.

The images were also pre-processed in order to identify and mask any foreground and background sources, such as stars and other galaxies, that lie in the projected vicinity of our objects. This way, subsequent analyses and measurements on the galaxy images are not contaminated by the light of interloping sources. Taking into consideration the small size of the data sample, we decided to individually inspect the images with DS9 and manually identify the position and extent of the interlopers. With this information, we then created a bad pixel mask image for each galaxy.

With these preparations complete, the next step consists in determining the basic properties of dwarf ETG sample, such as the brightness and geometry of the radial light profile of the galaxies. For this purpose, we make use of the isophotal analysis and construction tools available in an external package of IRAF, the Space Telescope Science Data Analysis System \citep[STSDAS;][]{stsdas:1994} package. To construct an accurate representation of the light profile, we use the IRAF \texttt{ellipse} task to fit each galaxy image with elliptical isophotes, while allowing their central coordinates, shape (ellipticity), and orientation (position angle) to change freely with galactocentric radius. Thus, any radial changes in the geometry of the galaxies are taken into account.

\begin{figure*}[ht]
\begin{center}
\gridline{\fig{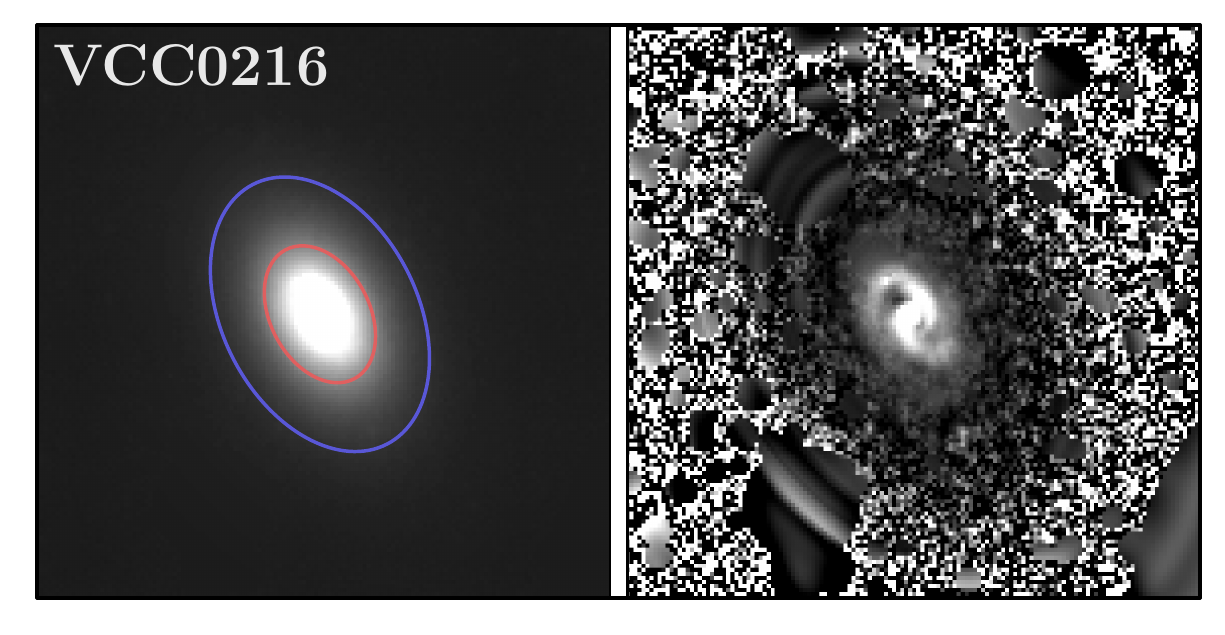}{0.334\textwidth}{}
          \hspace*{-10pt}
          \fig{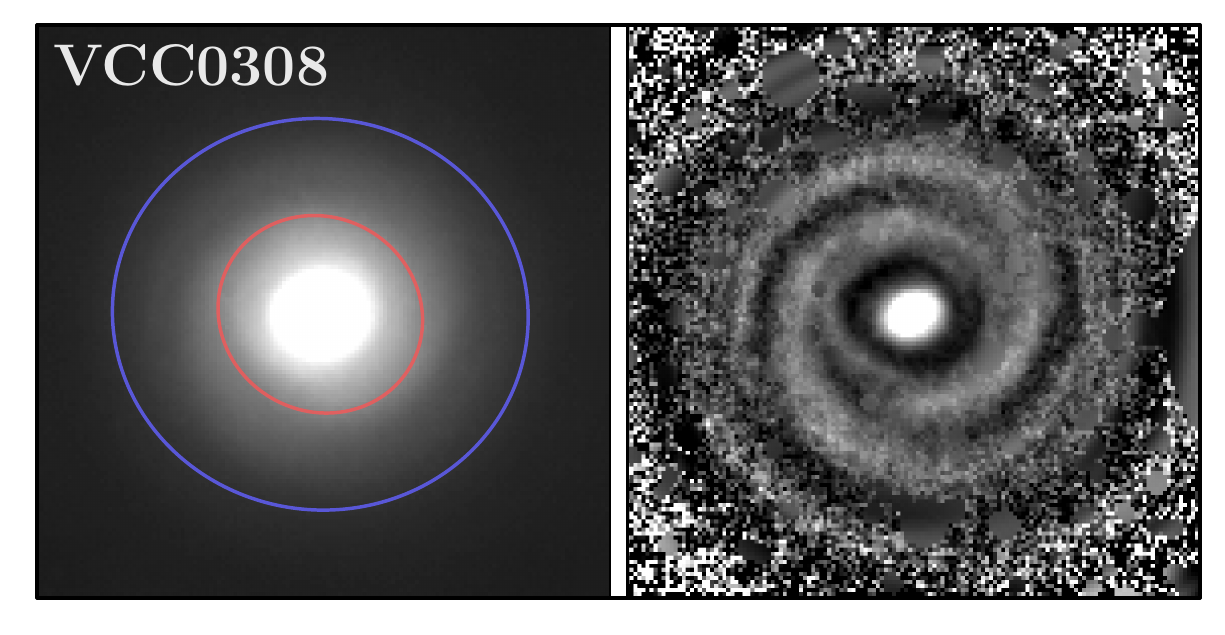}{0.334\textwidth}{}
          \hspace*{-10pt}
          \fig{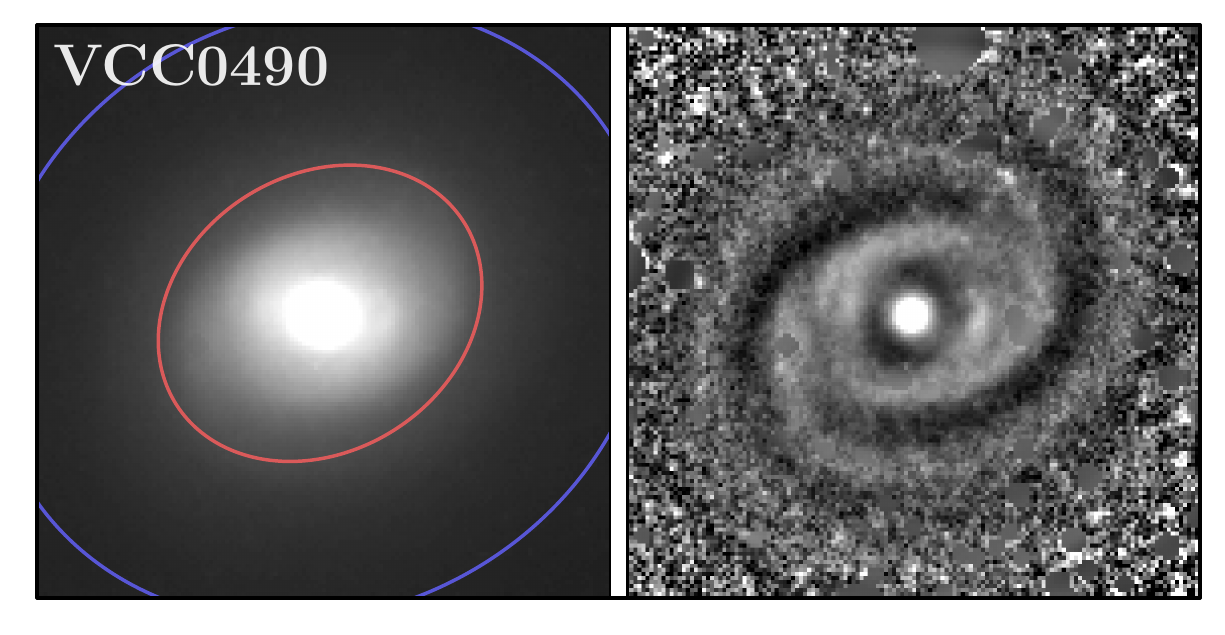}{0.334\textwidth}{}}
          \vspace*{-28.5pt}
\gridline{\fig{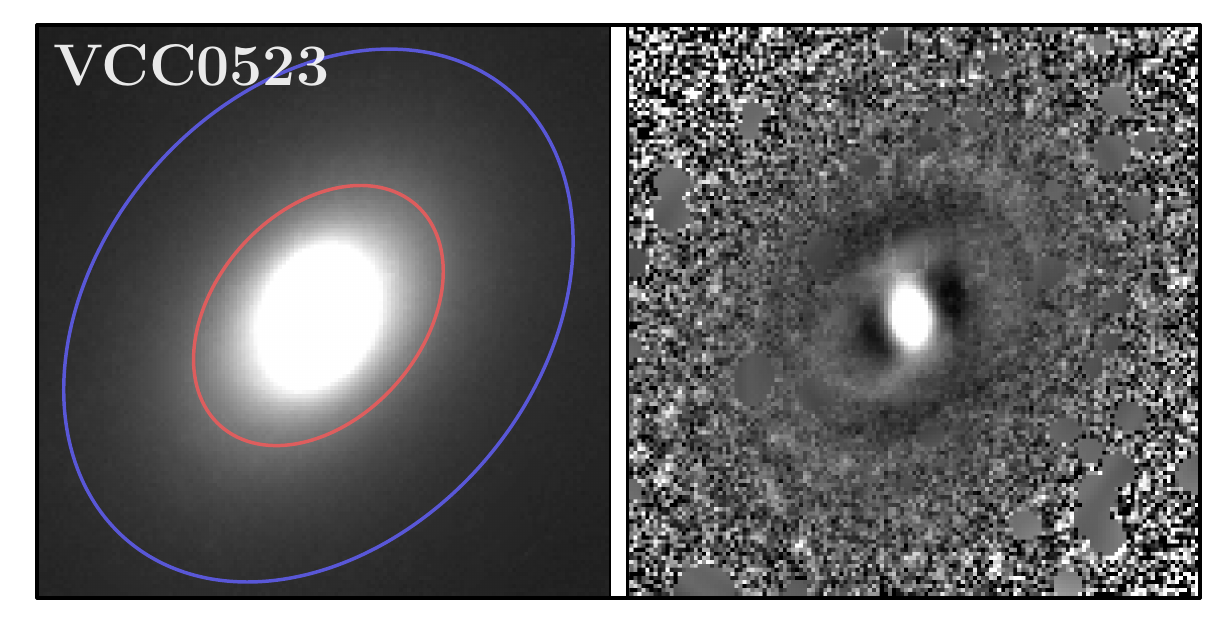}{0.334\textwidth}{}
          \hspace*{-10pt}
          \fig{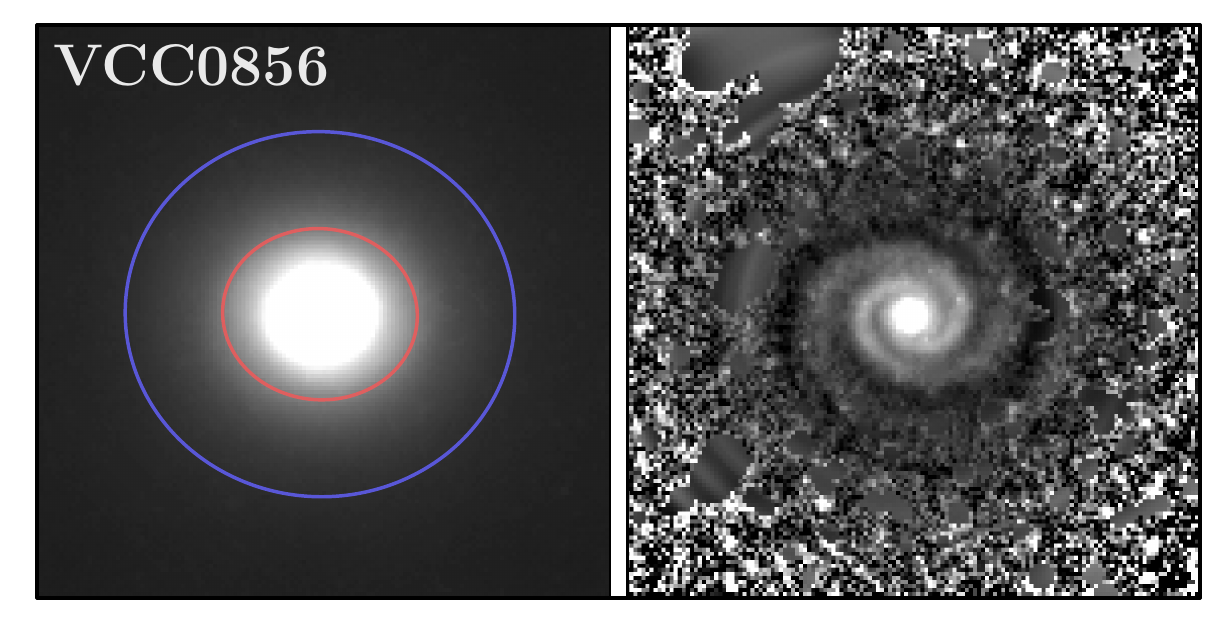}{0.334\textwidth}{}
          \hspace*{-10pt}
          \fig{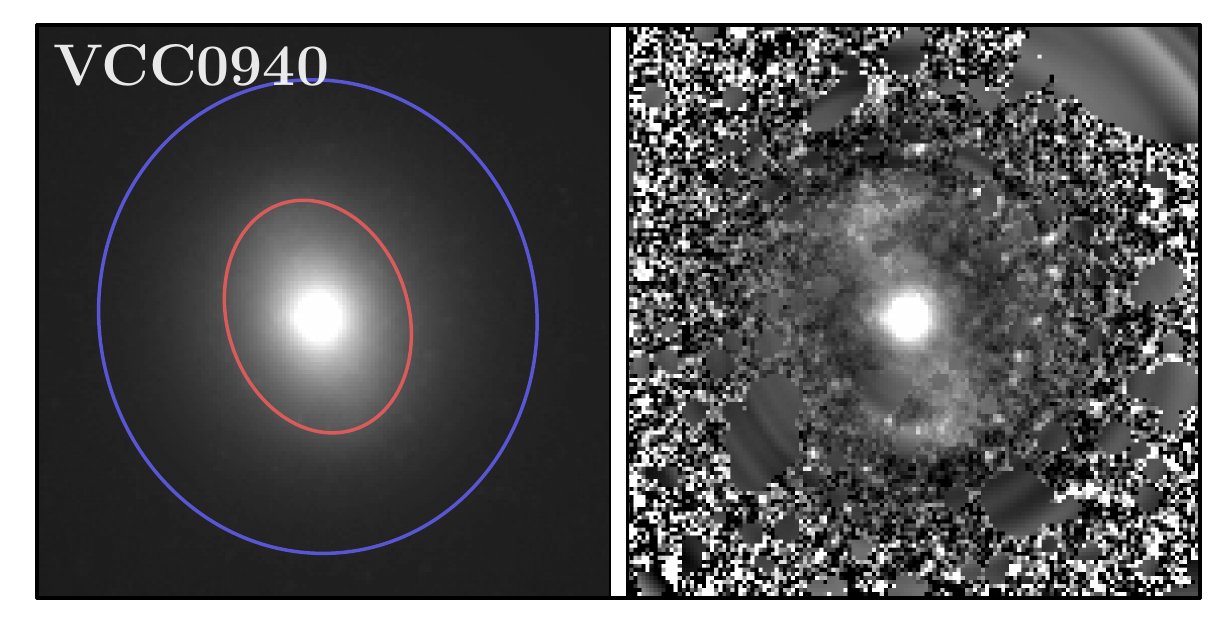}{0.334\textwidth}{}}
          \vspace*{-28.5pt}
\gridline{\fig{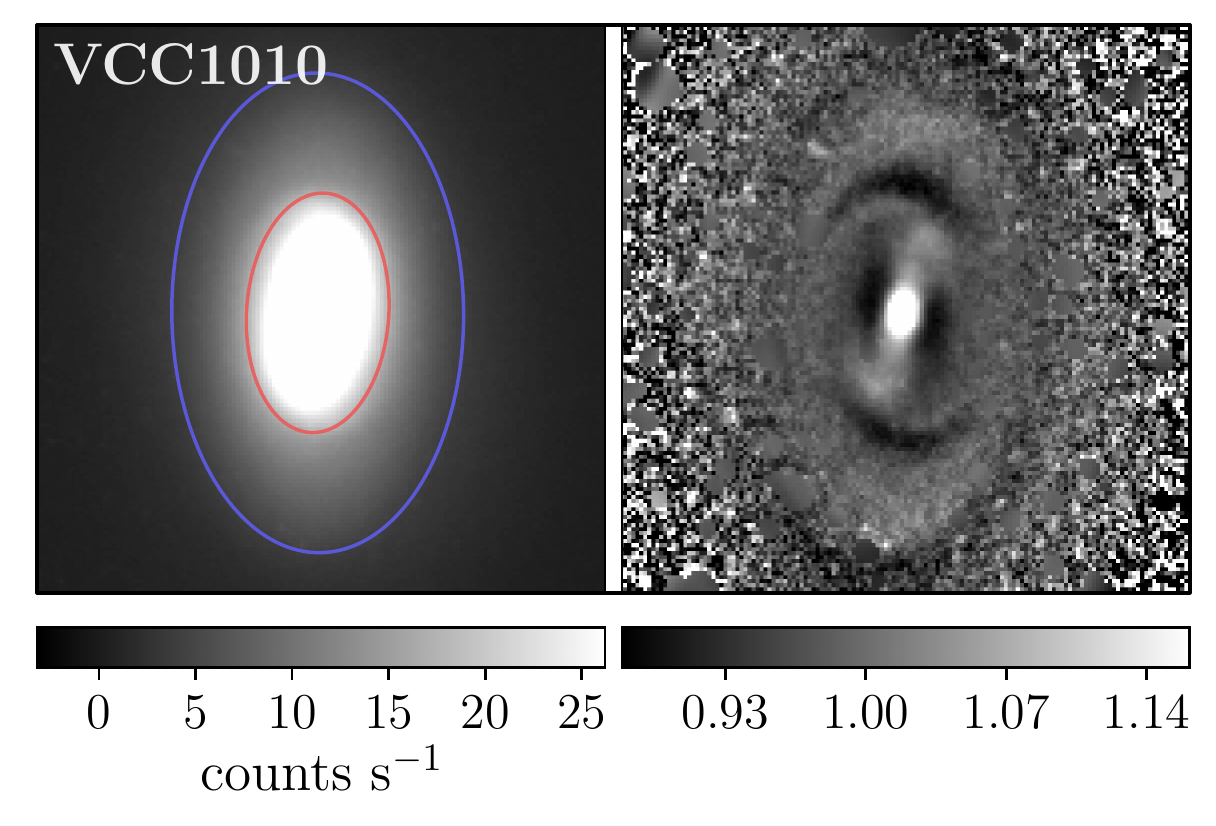}{0.334\textwidth}{}
          \hspace*{-10pt}
          \fig{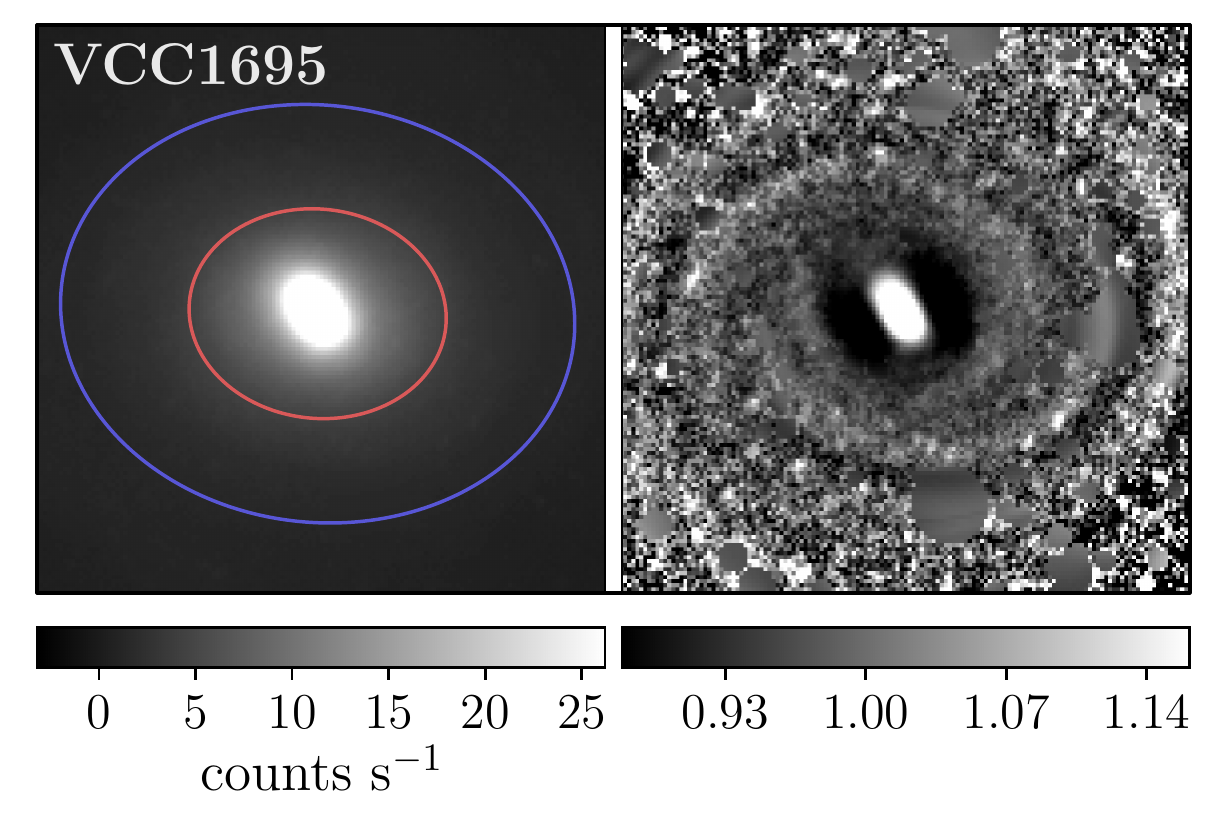}{0.334\textwidth}{}
          \hspace*{-10pt}
          \fig{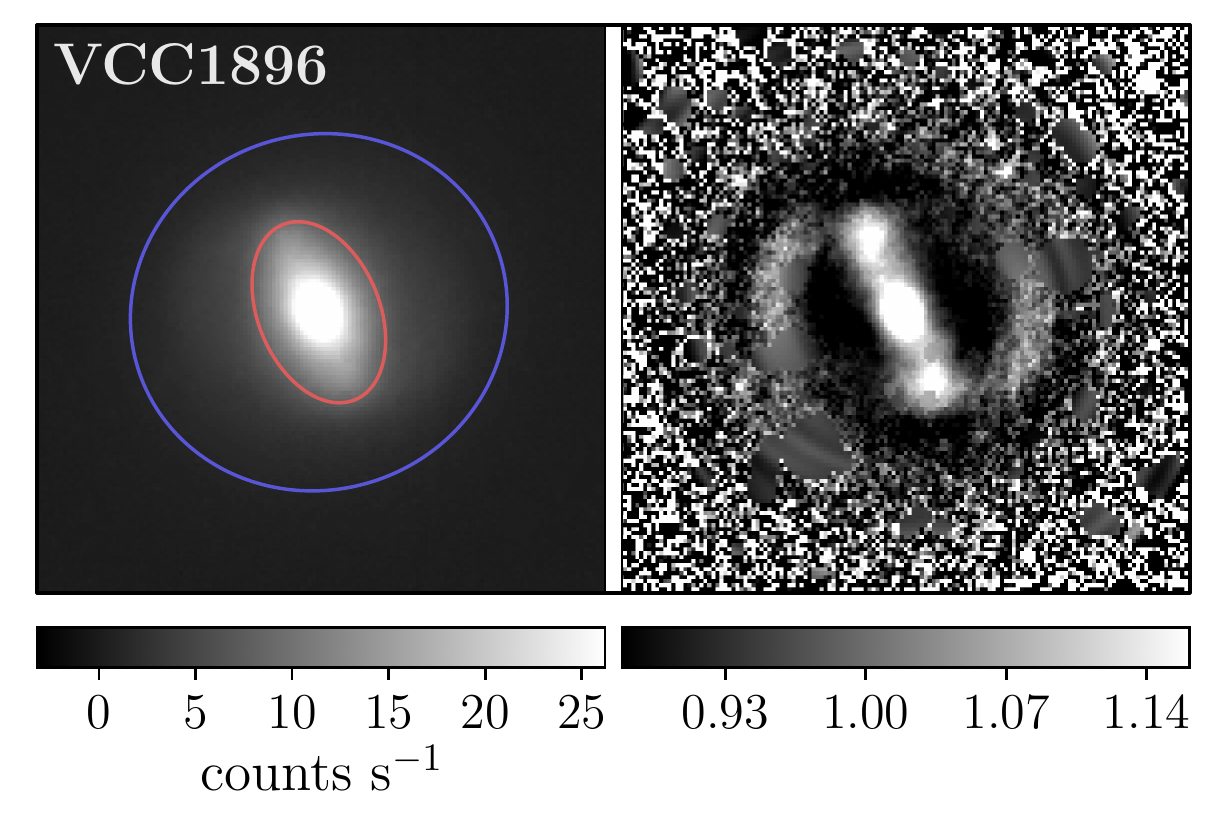}{0.334\textwidth}{}}
          \vspace*{-18pt}
\caption{{Original images and unsharp mask images of the dwarf ETG sample. Each image is labeled with the name of the dwarf galaxy it corresponds to, with its unsharp mask image displayed to the right. \myedit{The galaxy isophotes at one and two effective radii are overlaid on the original image as red and blue ellipses, respectively. Interloping sources have been masked out.} The columns share the same gray-scale bar; the original images are in units of counts s$^{-1}$, while the unsharp masks are unitless due to being image ratios. The images are $100\times100$ arcsec$^{2}$ in size. North is up, east is to the left.}\label{fig:data:imgs_gals_uns}}
\end{center}
\end{figure*}

With the information provided by the fit, we derive the photometric and structural properties of the data sample. These include the surface brightness, ellipticity, and position angle radial profiles, the total magnitude, the effective radius, and the central coordinates of the galaxies. First, we define the total galaxy area as the area enclosed by the last valid isophote, which corresponds to the isophote whose intensity is 2$\sigma$ above the background noise level. Then, from the flux enclosed by the last valid isophote, we compute the total flux of the galaxies and derive both the total magnitude and the effective radius of each galaxy, where the effective radius is defined as the radius that encloses half of the total galaxy light. For the galaxy center, we adopt the central coordinates of the isophote that has a semi-major axis length equal to 1.5 times the PSF FWHM. This isophote is chosen in order to avoid any effects of the PSF in the determination of the center. Additionally, we also use the IRAF \texttt{bmodel} task to translate the fit parameters into a galaxy model image. We then use this model image to create a galaxy image in which the interloping sources have been masked out and replaced with the corresponding value given by the model. As a result, we obtain a galaxy image that has a much cleaner, regular appearance. This is mainly for visualization purposes, as the regions marked on the bad pixel mask are still omitted during any kind of measurement or analysis. However, in the case that the galaxy image is subjected to smoothing, this prevents it from becoming contaminated by the blurred halos of bright interloping sources. Thus, as an example, the resulting unsharp mask image would be able to provide a clearer picture of any faint substructure that is present.

In order to visualize the data sample, we present the original galaxy images and their unsharp mask images in Figure \ref{fig:data:imgs_gals_uns}. For the creation of the unsharp masks, we divide the galaxy image with a smoothed-out version of itself. The galaxy image is smoothed out by using the IRAF \texttt{gauss} task to convolve it with an elliptical Gaussian kernel that matches the geometry (ellipticity and position angle) of the isophote at two effective radii, and which has a standard deviation of 4 arcsec. On the one hand, we choose the isophote at two effective radii to serve as a representation of the overall geometry of the main body of the galaxies, as at these radii they do not appear to be dominated by substructures. On the other hand, the choice of a standard deviation of 4 arcsec is tuned to match half of the average width of the disk features observed throughout the sample, in order for them to be smoothed out by just the necessary amount. This way, the smoothing process efficiently blurs out the disk substructure, while at the same time it mostly preserves and does not heavily alter the overall geometry of the galaxies.

We observe that the galaxy images are predominantly smooth in appearance, although in some cases it is possible to discern the presence of faint disk features towards the galaxy outskirts. In contrast, by removing most of the smooth light, the unsharp mask images reveal a rich variety of disk features \textendash such as bars, spiral arms, rings, and dumbbells\textendash\ that normally lie hidden in the dominant diffuse component of the galaxies.

These disk features constitute the galaxy substructures that we are interested in quantifying in this work. We would like to note that no previous attempts at quantification have been made on these data before. Therefore, this work complements and extends the previous studies that have analyzed this data set \citep{Lisker:2009c,Lisker:2009a,Lisker:2009b,Lisker:2012}.


\begin{deluxetable*}{cCCCCCCC}[ht]
\tablecaption{Central coordinates and photometric properties of the dwarf ETG sample.\label{tab:data:photo}}
\tablehead{
\colhead{Dwarf Galaxy} & \multicolumn2c{Central Coordinates} & \colhead{PSF FWHM} & \colhead{Depth $\mu_{r}$} & \colhead{$m_{r}$} & \colhead{$M_{r}$} & \colhead{$g-r$} \vspace*{1pt}\\
\cline{2-3}
 & \colhead{R.A.} & \colhead{Decl.} & \colhead{} & \colhead{} & \colhead{} & \colhead{} & \colhead{}\vspace*{-0.5pt}\\
  & \colhead{(J2000.0)} & \colhead{(J2000.0)} & \colhead{(arcsec)} & \colhead{(mag arcsec$^{-2}$)} & \colhead{(mag)} & \colhead{(mag)} & \colhead{(mag)}\\
\colhead{(1)} & \colhead{(2)} & \colhead{(3)} & \colhead{(4)} & \colhead{(5)} & \colhead{(6)} & \colhead{(7)} & \colhead{(8)} \vspace*{0.5pt}}
\startdata
VCC0216 & 12^{\text{h}}17^{\text{m}}01\pointsec\btwnpoint10 & +09^{\circ}24^{\prime}27\pointarcsec\btwnpoint11 & 1.13 & 27.94 & 14.42 & -16.67\pm0.15 & 0.603 \\
VCC0308 & 12^{\text{h}}18^{\text{m}}50\pointsec\btwnpoint91 & +07^{\circ}51^{\prime}42\pointarcsec\btwnpoint91 & 1.23 & 27.75 & 13.30 & -17.79\pm0.15 & 0.595 \\
VCC0490 & 12^{\text{h}}21^{\text{m}}38\pointsec\btwnpoint81 & +15^{\circ}44^{\prime}42\pointarcsec\btwnpoint30 & 1.27 & 27.91 & 13.06 & -18.03\pm0.15 & 0.648 \\
VCC0523 & 12^{\text{h}}22^{\text{m}}04\pointsec\btwnpoint13 & +12^{\circ}47^{\prime}14\pointarcsec\btwnpoint82 & 1.41 & 27.21 & 12.75 & -18.34\pm0.15 & 0.622 \\
VCC0856 & 12^{\text{h}}25^{\text{m}}57\pointsec\btwnpoint94 & +10^{\circ}03^{\prime}13\pointarcsec\btwnpoint55 & 1.29 & 27.89 & 13.47 & -17.62\pm0.15 & 0.636 \\
VCC0940 & 12^{\text{h}}26^{\text{m}}47\pointsec\btwnpoint07 & +12^{\circ}27^{\prime}14\pointarcsec\btwnpoint34 & 1.24 & 27.59 & 13.97 & -17.12\pm0.15 & 0.655 \\
VCC1010 & 12^{\text{h}}27^{\text{m}}27\pointsec\btwnpoint39 & +12^{\circ}17^{\prime}25\pointarcsec\btwnpoint08 & 1.21 & 27.66 & 12.87 & -18.22\pm0.15 & 0.693 \\
VCC1695 & 12^{\text{h}}36^{\text{m}}54\pointsec\btwnpoint86 & +12^{\circ}31^{\prime}12\pointarcsec\btwnpoint42 & 1.38 & 27.86 & 13.61 & -17.48\pm0.15 & 0.575 \\
VCC1896 & 12^{\text{h}}41^{\text{m}}54\pointsec\btwnpoint62 & +09^{\circ}35^{\prime}04\pointarcsec\btwnpoint58 & 1.27 & 27.79 & 14.25 & -16.84\pm0.15 & 0.618 \vspace*{1.5pt}\\
\enddata
\tablecomments{Col. (1): name of the dwarf galaxy. Cols. (2) and (3): right ascension and declination of the central coordinates in the International Celestial Reference System (ICRS). Col. (4): PSF FWHM of the \myedit{$3\times3$-binned} galaxy image. Col. (5): surface brightness depth in the $r$-band, at a $\text{S}/\text{N}=1$ and a scale of 0.71 arcsec pixel$^{-1}$. Col. (6): total apparent magnitude in the $r$-band. Col. (7): total absolute magnitude in the $r$-band, assuming a Virgo cluster distance modulus of $31.09 \pm 0.15$ mag \citep{Blakeslee:2009}. Col. (8): $g-r$ color integrated within two effective radii \citep{Janz:2008,Janz:2009}.}
\end{deluxetable*}

We now proceed to describe the photometric and structural properties of the data sample. First, we provide the central coordinates and photometric properties of the galaxies, which are reported in Table \ref{tab:data:photo}. The galaxies are intrinsically faint, with an absolute magnitude in the SDSS $r$-band equivalent that lies in the range $-18.4 < M_{r} < -16.6$ mag; thus spanning \myedit{almost} two magnitudes. Although we only have the white filter imaging to work with, and thus lack color information, we provide the $g-r$ colors of the galaxies as a reference. These color measurements were performed by \citet{Janz:2008,Janz:2009} based on calibrated images of \citet{Lisker:2007,Lisker:2008} from the SDSS Data Release 5 \citep[DR5;][]{Adelman-McCarthy:2007}. The $g-r$ color of the galaxies, integrated up to two effective radii, is predominantly red and very uniform across the sample, spanning the range $0.58 \leq g-r \leq 0.69$ mag. Therefore, the data sample is characterized by having a low intrinsic brightness and being red in color.


Finally, the structural properties of the data sample are reported in Table \ref{tab:data:gal_params}. We provide the effective radius, and both the ellipticity and position angle at one and two effective radii of the galaxies. For the development of the procedure that will be used to quantify the galaxy disk substructure, knowledge of these parameter values is necessary. We observe that, in contrast to the homogeneity of the photometric properties, the structural properties of the galaxies are much more diverse. While there are some cases in which the ellipticity and position angle of the galaxy isophotes are similar at one and two effective radii, there are also cases in which their values change substantially. \myedit{This can be visualized in Figure \ref{fig:data:imgs_gals_uns}, where the isophotes at one and two effective radii have been overlaid on the galaxy images.} Due to this variety, the quantification method we have developed, described in the next Section \ref{sec:method}, is adapted to work properly in the case of both geometrically simple and complex radial light profiles.


\begin{deluxetable*}{cCCC@{\hspace{0.5em}}c@{\hspace{0.5em}}DD}[ht]
\tablecaption{Structural parameters of the dwarf ETG sample.\label{tab:data:gal_params}}
\tablehead{
\colhead{Dwarf Galaxy} & \colhead{Effective Radius} & \multicolumn2c{Ellipticity} & & \multicolumn4c{Position Angle} \vspace*{1pt}\\
\cline{3-4}\cline{6-9}
 & & \colhead{At $1 R_{e}$} & \colhead{At $2 R_{e}$} & & \multicolumn2c{At $1 R_{e}$} & \multicolumn2c{At $2 R_{e}$}\\
 & \colhead{(arcsec)} & & & & \multicolumn2c{(deg)} & \multicolumn2c{(deg)}\\
\colhead{(1)} & \colhead{(2)} & \colhead{(3)} & \colhead{(4)} & & \multicolumn2c{(5)} & \multicolumn2c{(6)} \vspace*{0.5pt}}
\decimals
\startdata
VCC0216 & 10.63\pm0.01 & 0.324\pm0.002 & 0.333\pm0.002 & & 28.3\pm0.1 & 27.4\pm0.2 \\
VCC0308 & 17.77\pm0.04 & 0.068\pm0.004 & 0.060\pm0.003 & & 59.9\pm1.9 & 83.0\pm1.4 \\
VCC0490 & 27.16\pm0.05 & 0.188\pm0.003 & 0.122\pm0.006 & & 122.3\pm0.5 & 118.0\pm1.6 \\
VCC0523 & 21.88\pm0.02 & 0.298\pm0.001 & 0.252\pm0.004 & & 138.4\pm0.1 & 139.6\pm0.5 \\
VCC0856 & 16.16\pm0.01 & 0.122\pm0.002 & 0.063\pm0.005 & & 85.6\pm0.4 & 86.3\pm2.2 \\
VCC0940 & 18.35\pm0.03 & 0.236\pm0.003 & 0.077\pm0.009 & & 17.6\pm0.4 & 8.5\pm3.5 \\
VCC1010 & 16.49\pm0.03 & 0.407\pm0.002 & 0.392\pm0.002 & & 176.3\pm0.2 & 0.5\pm0.2 \\
VCC1695 & 20.73\pm0.03 & 0.188\pm0.002 & 0.191\pm0.008 & & 83.8\pm0.4 & 83.5\pm1.2 \\
VCC1896 & 13.54\pm0.05 & 0.358\pm0.005 & 0.062\pm0.006 & & 22.5\pm0.5 & 106.8\pm3.0 \vspace*{1.5pt}\\
\enddata
\tablecomments{Col. (1): name of the dwarf galaxy. Col. (2): circularized effective radius, defined as $R_{e} = \sqrt{ab}$, where $a$ and $b$ are the semi-major and semi-minor axis lengths of the elliptical isophote, respectively. Cols. (3) and (4): ellipticity at one and two effective radii, respectively. The ellipticity is defined as $e = 1 - ba^{-1}$. Cols. (5) and (6): position angle at one and two effective radii, respectively. The position angle is measured counterclockwise from the +\textit{y}-axis (north towards east of the images).}
\end{deluxetable*}

\section{Residual method}\label{sec:method}

Our ``residual method'' consists of a newly developed procedure that aims to extract the disk features, such as bars and spiral arms, present in a galaxy image. Its development, subsequent testing, and fine tuning were carried out based on its application to the dwarf ETG sample presented in this work. Thus, we tested its intended functionality and found it to work reliably in the case that faint disk substructure is embedded in the more homogeneous, and much brighter, main body of a galaxy.

The residual method is an iterative procedure that gradually separates a galaxy image into two distinct components. On the one hand, it produces a galaxy model image that contains the dominant, diffuse component of the galaxy. On the other hand, it produces a galaxy residual image that contains the secondary disk component of the galaxy. Initially, the galaxy model image is partially contaminated by light coming from the galaxy's underlying disk features. Through an iterative loop, this extra light is progressively shifted to the galaxy residual image. The iterations end once the level of pollution in the model is minimized, and the majority of the substructure light is contained in the residual map. Thus, the galaxy residual image can then be used to quantify the relative contribution of the disk component to the total light of the galaxy. 

\myedit{It is important to note that throughout this work, the ``disk component'' is defined as the non-smooth, excess light contained in the disk substructures of a galaxy. It does not correspond to the total amount of light that may be contained in a physically thin embedded disk, which may also have an additional thin smooth component of its own. The residual method separates smooth (axisymmetric) light from non-smooth (non-axisymmetric) light, and thus cannot distinguish between different sources of smooth light (e.g., the diffuse light of a thin disk from that of a thick disk).}

The method makes use of IRAF built-in tasks, and of the isophotal analysis and construction tools available in the STSDAS \texttt{isophote} sub-package. In the following Section \ref{subsec:method:steps}, we provide a step-by-step description of the residual method, given as a recipe to follow. Then, in Section \ref{subsec:method:config}, we address the free parameters that must be set when running the residual method, and explain the specific configuration we adopted when applying the method to the dwarf ETG sample.

\subsection{Steps of the Residual Method}\label{subsec:method:steps}

\begin{figure*}[ht]
\begin{center}
\includegraphics[width=0.8\textwidth]{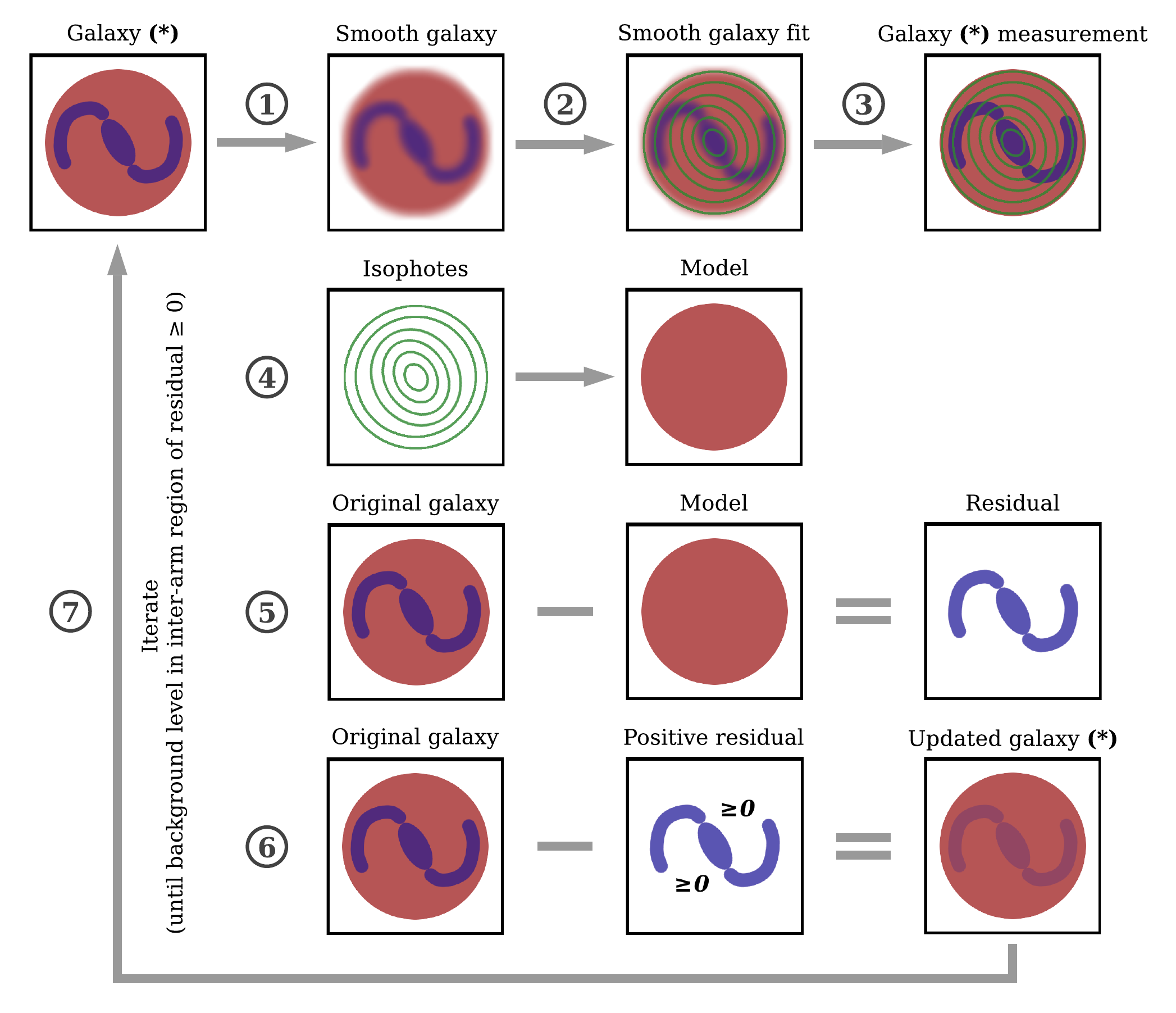}
\caption{Flowchart of the residual method. Steps 1 through 7 are indicated, as described in the text. The diffuse component of the galaxy is illustrated in red and the disk component in blue. The isophotal fit to the galaxy is illustrated by green ellipses. The updated, \myedit{positive-residual}-subtracted galaxy image resulting from the procedure is marked with an asterisk (*), and replaces the galaxy images that are similarly marked during the iterative loop.\label{fig:method:flowchart}}
\end{center}
\end{figure*}

We now proceed to describe the steps that comprise the residual method. First, each step provides an explanation on what it aims to achieve, and why it is necessary. This is followed by a description of the specific IRAF tasks and parameter settings that are utilized for this purpose. As a helpful reference, we provide a flowchart that illustrates the steps of the residual method in Figure \ref{fig:method:flowchart}.\\

\noindent \textbf{Step 1: Smooth out the galaxy image.} First, the objective is to smooth out the features of the disk component of the galaxy while at the same time preserving the main geometry of the diffuse component. By smoothing the galaxy image by a specifically tuned amount, the disk substructure embedded in the galaxy becomes washed out and less defined in appearance. Thus, there is a reduction in the relative contribution of these features to the brightness, shape, and orientation of the radial light profile of the galaxy.

Using the IRAF \texttt{gauss} task, the galaxy image is smoothed out by convolving it with an elliptical Gaussian kernel. The smoothing kernel has a shape and orientation that matches the ellipticity and position angle of the original galaxy image at two effective radii. When inspecting the unsharp mask images of the galaxies (see Figure \ref{fig:data:imgs_gals_uns}), the disk features appear faint and not dominant at this galactocentric distance. Thus, the two effective radii isophote is chosen to represent the geometry of the main body of the galaxy and is the geometry adopted for the smoothing kernel. Additionally, for the choice of kernel size, the observed thickness or width of the disk features of the galaxy is taken into account. The standard deviation of the Gaussian kernel is then tuned to match half of the average width of the disk features. \myedit{In the case of our dwarf ETG sample, the disk features have average half-widths ranging between $2.5-6.0$ arcsec.}\\

\noindent \textbf{Step 2: Fit the smoothed-out galaxy image.} By fitting the radial light profile of the smoothed-out galaxy image, we construct a representation of the galaxy that is less affected by the radial changes in brightness, shape, and orientation that may be driven by the underlying disk substructure. \myedit{In an ideal case, the influence of the disk features would be removed completely in Step 1, so the radial fit of the smoothed-out galaxy image would describe the geometry of the diffuse component. However, in reality, the disk features are still partially contaminating the light of the smoothed-out galaxy. Therefore,} the resulting fit describes a diffuse component that is still slightly influenced by the geometry of the disk component.

Using the IRAF \texttt{ellipse} task, the smoothed-out galaxy image is fit with concentric elliptical isophotes. The fit performed allows both the ellipticity and position angle of the isophotes to change freely with galactocentric radius, while their central coordinates are kept fixed. By adopting this configuration, the smoothed-out galaxy image is described as accurately as possible, as the fit accounts for any radial changes in the shape and orientation of the light profile. The central coordinates, however, are kept fixed to ensure the convergence of the \texttt{ellipse} task even in the regions of the galaxy that have a highly flattened radial light gradient, which can arise as a result of the smoothing procedure in Step 1. The adopted galaxy center is extracted from a free fit previously performed on the original galaxy image (see Section \ref{sec:data}). It corresponds to the central coordinates of the isophote that has a semi-major axis equal to 1.5 times the PSF FWHM, chosen in order to avoid any effects of the PSF in the determination of the center. 

\myedit{Additionally, there are other parameters of the \texttt{ellipse} task that influence the result of the fit. In order to filter out and flag outlier points from the average along each isophote, we have adopted three-sigma clipping during the fit ($\texttt{usclip}=\texttt{lsclip}=3$), which iterates five times ($\texttt{nclip}=5$). We also allow up to 50\% of flagged points along a valid isophote ($\texttt{fflag}=0.5$).}\\

\noindent \textbf{Step 3: Measure the galaxy image by imposing the isophotes of the smoothed-out galaxy image.} Inherent to the process of smoothing out the galaxy image in Step 1, light from the galaxy is redistributed throughout the image. Therefore, when fitting the smoothed-out galaxy image in Step 2, the extracted isophotes do not reflect the actual intensity profile of the galaxy. To obtain the true, \myedit{average} intensity at each isophote, we return to the galaxy image itself and superimpose these \myedit{isophotal} contours on the galaxy to measure the true, \myedit{average} underlying brightness. This way, we obtain a representation of the radial light profile of the galaxy that reflects the \myedit{actual} brightness of the galaxy image while following the shape and orientation of the smoothed-out galaxy image.

Using the IRAF \texttt{ellipse} task in no-fit, photometry-only mode, we measure the \myedit{average} radial brightness of the galaxy image by imposing the geometry of the isophotes of the smoothed-out galaxy image. This is achieved by setting the \texttt{inellip} parameter of the task, and providing the \texttt{ellipse} output table obtained from fitting the smoothed-out galaxy in Step 2.\\

\noindent \textbf{Step 4: Construct the galaxy model image, representing the diffuse component.} From the previous step, we now have a description of the light profile of the galaxy in which the impact of the underlying disk substructure has been minimized. In other words, the extracted light profile aims at describing the diffuse component of the galaxy. In order to construct a \myedit{two-dimensional} representation of the diffuse component, \myedit{we use the information we have extracted of the radial shape, orientation, and brightness of the light profile to build a model image.}

Using the IRAF \texttt{bmodel} task, we create a galaxy model image that represents the diffuse component of the galaxy. This is achieved by providing the task with the \texttt{ellipse} output table obtained in Step 3.\\

\noindent \textbf{Step 5: Construct the galaxy residual image, representing the disk component.} If the contribution of the diffuse component is removed from the galaxy image, then what is left behind constitutes the disk component. That is, any substructure features embedded in the galaxy, such as spiral arms and bars, emerge as residual light when subtracting out the diffuse light of the galaxy.

Using the IRAF \texttt{imarith} task, the galaxy model image from Step 4 is subtracted from the original galaxy image. As a result, we obtain a galaxy residual image that contains the disk component of the galaxy.\\

\noindent \textbf{Step 6: Construct the updated, \myedit{positive-residual}-subtracted version of the galaxy image.} Conceptually, the galaxy model image (Step 4) should only contain the diffuse light of the galaxy, in order for the galaxy residual image (Step 5) to contain the remaining light coming from substructure features. However, the separation into these two distinct components is not achieved in one go. Inevitably, the galaxy model image will be partially contaminated with light coming from the disk features, as it is not possible to completely remove their influence when smoothing out the galaxy image in Step 1. Consequently, the galaxy model tends to be slightly brighter than it should be, resulting in a galaxy residual that is slightly fainter and that suffers from over-subtraction. In other words, the inter-arm region in the galaxy residual, by being devoid of disk features, is dominated by negative values.

Therefore, the extra light that is coming from disk features and that is contaminating the galaxy model has to be shifted instead into the galaxy residual, where it belongs. To achieve this, we construct an updated version of the galaxy image in which \myedit{all of} the positive light from the galaxy residual image has been subtracted \myedit{out}. This new galaxy image, which is less driven by the influence of the disk features, can be then once again subjected to Steps 1 through 5.

First, using the IRAF \texttt{imreplace} task, the negative values in the inter-arm region of the galaxy residual image are set to zero. This way, we get rid of the effects of over-subtraction and obtain a galaxy residual image that has only positive values along its disk features. Then, using the IRAF \texttt{imarith} task, this positive-only galaxy residual image is subtracted from the original galaxy image. As a result, we obtain an updated, \myedit{positive-residual}-subtracted version of the galaxy image that can then be used again in Step 1.\\

\noindent \textbf{Step 7: Iterate until the inter-arm region in the galaxy residual image reaches a level of zero.} By entering an iterative loop of Steps 1 through 6, the galaxy becomes gradually separated into a diffuse and a disk component, contained in the galaxy model (Step 4) and galaxy residual (Step 5) images, respectively. This iterative procedure is carried on until the galaxy decomposition is complete. To know when we have reached this point in the loop, we construct a stopping criterion for the iterations by monitoring the inter-arm region of the galaxy residual image. \myedit{We define the ``inter-arm region'' as all the pixels that are negative in the initial version of the residual image (i.e., in the first residual image that is created before initiating the iterative loop), and that lie within the two effective radii isophote of the galaxy.}

Initially dominated by negative values, the inter-arm region becomes each time less negative with successive iterations, gradually alleviating the effects of over-subtraction. \myedit{As our aim is to measure the excess light contained in disk features, the iterations should continue until the inter-arm region reaches a typical statistical value that is on par with the background level. While the images in our sample have been background subtracted, they also contain noise, so their background consists in positive and negative values that fluctuate around zero. We have chosen the median as the statistical function to estimate the central value of the inter-arm region, as it is robust against possible outliers.} Thus, the final, stopping iteration is reached once the inter-arm region of the galaxy residual first reaches a median value equal or bigger than zero. We note that, if the iterations were to continue after this point, the galaxy residual would start suffering instead from under-subtraction. That is, too much light would begin to be displaced from the galaxy model to the galaxy residual, making the latter artificially brighter. Consequently, the stopping criterion for the iterations is designed to prevent both over- and under-subtraction in the galaxy residual image.

\begin{figure}[ht]
\includegraphics[width=\columnwidth]{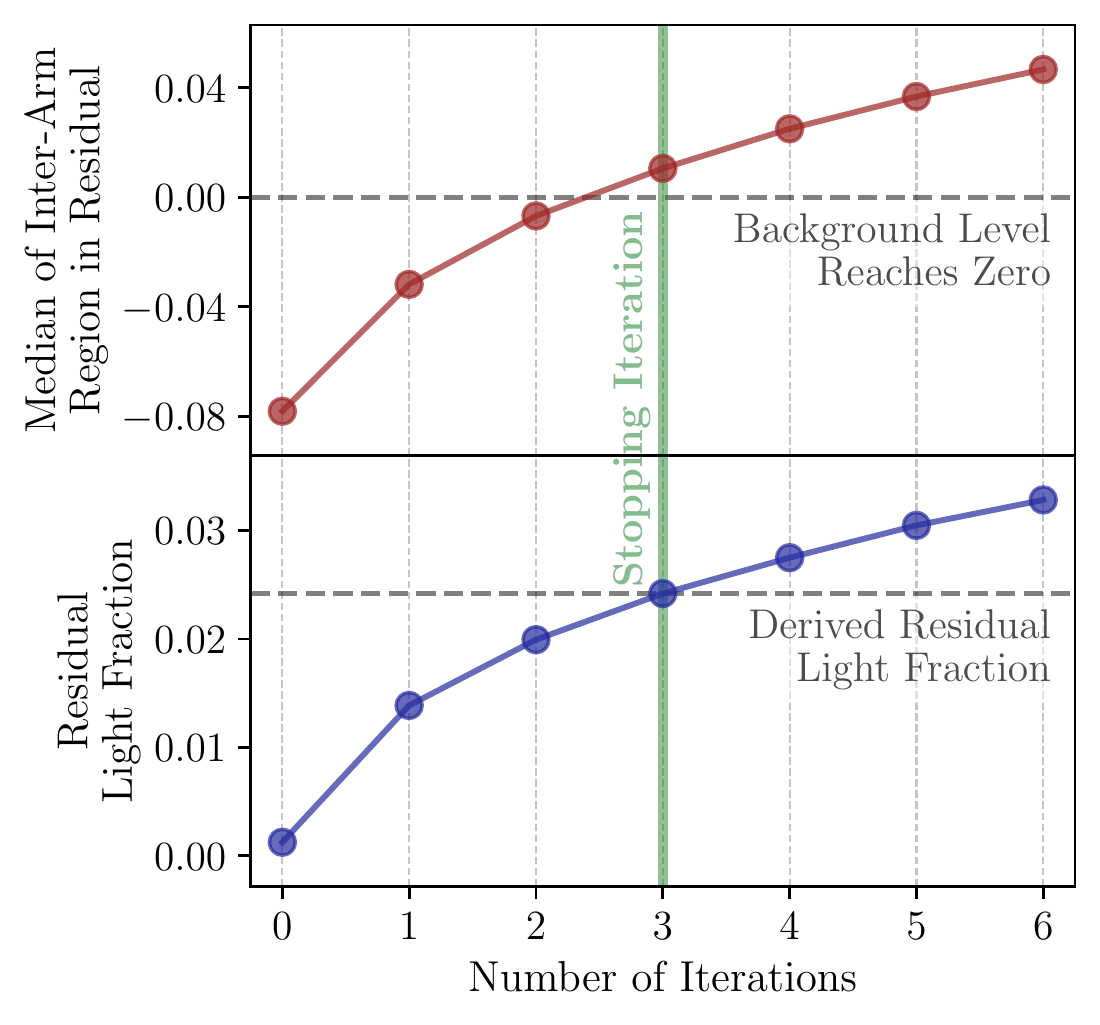}
\caption{Schematic representation of the iterative procedure and stopping criterion implemented in the residual method. The data points correspond to empirical values obtained when applying the residual method to the dwarf galaxy VCC0308. \textit{Top panel:} the median value of the inter-arm region in the galaxy residual image vs. the number of iterations, represented by a red curve. The specific iteration at which the background level in this region becomes equal or bigger than zero is marked with a vertical green line, and corresponds to the stopping iteration. \textit{Bottom panel:} the measured residual light fraction of the galaxy vs. the number of iterations, represented by a blue curve. The accepted value for the residual light fraction is derived from the galaxy residual image at the stopping iteration, as shown by a horizontal dashed line.\label{fig:method:schematic_its}}
\end{figure}

With its background at a zero level, the final, optimized version of the galaxy residual image can be then used to accurately quantify the amount of light contained in the disk features. For reference, a schematic representation of the iterative procedure and the stopping criterion is provided in Figure \ref{fig:method:schematic_its}.
\newline\newline\newline\newline

\subsection{Parameter Configuration of the Residual Method}\label{subsec:method:config}

When running the residual method, we adopt a parameter setup that is specifically tuned to the analysis of the dwarf ETG sample presented in Section \ref{sec:data}. This configuration needs to be tested and adjusted on a case-by-case basis, in accordance to the properties of the particular data set that the method is going to be applied to. This includes properties both of the images and of the dwarf galaxies themselves. Thus, factors such as the S/N, the seeing and PSF, and the resolution of the images should be taken into account. In regard to the galaxies, the average width and relative strength of their disk features will also influence the parameter configuration. For this reason, the parameter values that are described in this section should only be used as a guideline.

The first parameter to be configured corresponds to the smoothing kernel size. In Step 1 of the residual method, the galaxy image is smoothed by convolving it with an elliptical Gaussian kernel. As the objective is to blur out the disk substructure \myedit{light while at the same time preserving the overall geometry of the diffuse light}, the standard deviation of the Gaussian smoothing kernel is adjusted to match half of the average width of the disk features. This can be understood as follows. On the one hand, if the kernel size is much smaller than the thickness of the bars and spiral arms, then their brightness and appearance are not significantly altered during the smoothing process. On the other hand, if the kernel size is instead much bigger, then not only the disk features are blurred out, but also the main body of the galaxy becomes strongly affected. Consequently, there is a lower and an upper limit on the optimal kernel sizes that can be used in the smoothing procedure. For the dwarf ETG sample, the optimal range corresponds to a Gaussian standard deviation between $2.5-6.0$ arcsec in size, as this range encompasses the average \myedit{half-width} of the disk features throughout the whole sample. In order to subject the whole data set to the same treatment, we apply this range of smoothing kernel sizes to all of the galaxies. Additionally, through testing, we found that once the galaxy is smoothed one time, the majority of the disk features already become successfully blurred out, so it is not necessary to continue smoothing them by the same amount in subsequent iterations. Therefore, to reduce the amount of smoothing to the minimum required, and thus to help preserve the geometry of the diffuse component, the choice of the kernel size is decreased to a half once the galaxy image enters the iterative loop of the method. In other words, if the galaxy image is being initially smoothed out by a kernel of size $X \in [2.5,6.0]$ arcsec, this size is reduced to just $X/2$ arcsec from the first iteration onwards.

The second parameter to be configured corresponds to the sampling step size. In Step 2 of the residual method, the smoothed-out galaxy image is fit with concentric elliptical isophotes. By adjusting the separation between successive isophotes, it is possible to sample the image at a greater or a lower frequency, corresponding to smaller or larger step sizes, respectively. On the one hand, if the step sizes are too small, the fitting task does not have enough data points to sample the image and construct the isophotes. On the other hand, if the step sizes are instead too big, the resulting isophotes are unable to accurately capture any subtle radial changes in the brightness and geometry of the galaxy. Therefore, once again depending on the properties of the data set, there is an optimal range of sampling step sizes. For the dwarf ETG sample, we opt for a geometric growth of the step size instead of a linear growth, as it provides the fitting procedure with more data points to construct the isophotes towards the fainter, lower S/N regions in the galaxy outskirts. We determine that a growth rate of the step size length in the range of $5-10$\% is optimal for our sample. \myedit{In terms of the overall \texttt{ellipse} sampling setup, we always implement three-sigma clipping along each isophote in order to filter out outlier values, as detailed in Step 2.}

\begin{figure*}[ht]
\begin{center}
\includegraphics[width=0.7\textwidth]{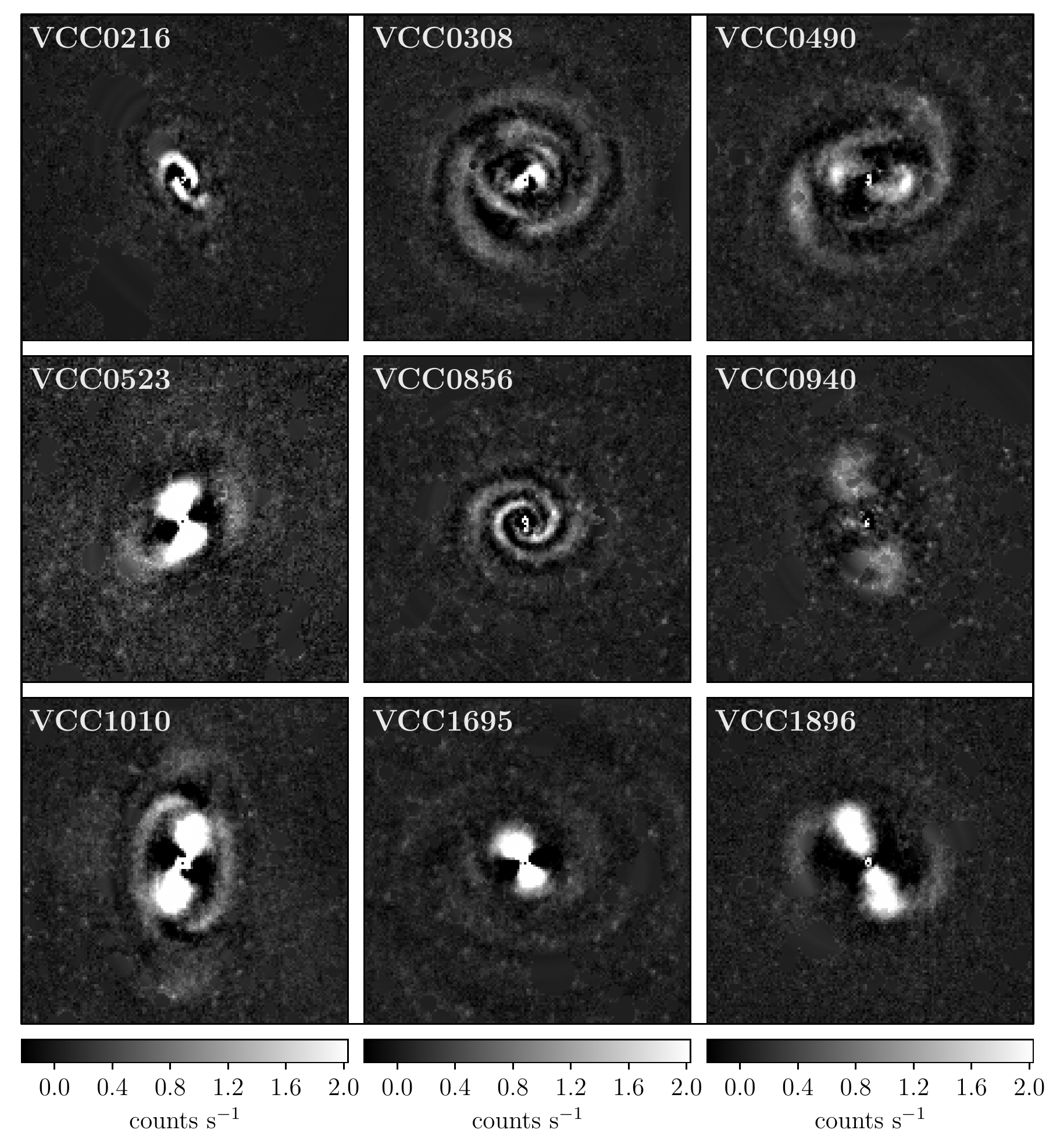}
\caption{Galaxy residual images obtained when applying the residual method to the dwarf ETG sample. The specific parameter setup used to obtain these images corresponds to a Gaussian standard deviation of 4 arcsec for the smoothing kernel size and a growth rate of 7\% for the sampling step size. Each residual image is labeled with the name of the dwarf galaxy it corresponds to. All images share the same gray-scale bar in units of counts s$^{-1}$, and are $100\times100$ arcsec$^{2}$ in size. North is up, east is to the left.\label{fig:results:imgs_res}}
\end{center}
\end{figure*}

After thorough testing, we can confirm that the residual method is quite insensitive to the input parameters when they are chosen within reason, such as we have done for our dwarf ETG sample. To implement the aforementioned reasonable ranges of smoothing kernel sizes and sampling step sizes, the residual method is run several times on each galaxy image. Each run is set up with a specific combination \myedit{pair} of smoothing and sampling parameter values. Depending on this configuration, the number of iterations needed in the procedure ranges between two to five iterations, with a median of three iterations. Thus, by exploring the smoothing and sampling parameter spaces and \myedit{by applying the stopping criterion}, we obtain as a result a set of residual images for each galaxy that only vary slightly with respect to one another. These galaxy residual images are then used to quantify the strength of the disk substructure and its related uncertainties, as described in Section \ref{sec:results}.

\section{Results}\label{sec:results}

We now present the results obtained from the application of the residual method to the dwarf ETG sample. The method allows us to isolate the light of the diffuse component of the galaxy in a galaxy model image, while the remaining light coming from the disk component is contained in a galaxy residual image. Thus, any underlying disk substructure can be visualized through the inspection of the galaxy residual images, which are shown in Figure \ref{fig:results:imgs_res}. We observe a wide diversity of disk features that, based on a purely qualitative assessment, include bars (e.g., VCC1695, VCC1896), spiral arms (e.g., VCC0308, VCC0856), rings (e.g., VCC0490, VCC1010), and dumbbells (e.g., VCC1010). These features also appear to have different brightness distributions, being in some cases brighter and in others fainter. From a direct comparison with the unsharp mask images presented in Figure \ref{fig:data:imgs_gals_uns}, it is possible to confirm that all the disk features visible in the unsharp masks are captured successfully by the residual method in the residual images.


To quantify the brightness of the disk component of the galaxy, we measure the amount of light contained in the galaxy residual image compared to the amount of light contained in the original galaxy image. The quantity we are interested in corresponds to the ratio of the residual-to-total light, which we define as the ``residual light fraction''. In the case of the dwarf ETG sample, we decide to measure the residual light fraction within both one and two effective radii. We note that these one and two effective radii measurements do not assume circularized radii, but instead match the ellipticity and position angle of the one and two effective radii isophotes of the galaxies (as reported in Table \ref{tab:data:gal_params} \myedit{and shown in Figure \ref{fig:data:imgs_gals_uns}}). To compute the residual light fraction, the light contained in each image, original and residual, is integrated up to the same given elliptical isophote. \myedit{These sums include all enclosed pixel values, regardless of their sign. The ratio of these two sums is then calculated.}
  
\begin{deluxetable}{cCC}[ht]
\tablecaption{Residual light fractions of the dwarf ETG sample.\label{tab:results:res}}
\tablehead{
\colhead{Dwarf Galaxy} & \multicolumn2c{Residual Light Fraction} \vspace*{1pt}\\
\cline{2-3}
 & \colhead{Within $1 R_{e}$} & \colhead{Within $2 R_{e}$} \vspace*{-1pt}\\
\colhead{(1)} & \colhead{(2)} & \colhead{(3)} \vspace*{0.5pt}}
\startdata
VCC0216 & 0.033^{+0.002}_{-0.003} & 0.029^{+0.004}_{-0.002} \\
VCC0308 & 0.017^{+0.002}_{-0.001} & 0.024^{+0.002}_{-0.002} \\
VCC0490 & 0.029^{+0.004}_{-0.004} & 0.036^{+0.004}_{-0.003} \\
VCC0523 & 0.040^{+0.005}_{-0.006} & 0.042^{+0.003}_{-0.004} \\
VCC0856 & 0.017^{+0.001}_{-0.001} & 0.022^{+0.004}_{-0.002} \\
VCC0940 & 0.024^{+0.003}_{-0.003} & 0.035^{+0.003}_{-0.003} \\
VCC1010 & 0.031^{+0.005}_{-0.006} & 0.030^{+0.004}_{-0.005} \\
VCC1695 & 0.058^{+0.008}_{-0.010} & 0.054^{+0.005}_{-0.007} \\
VCC1896 & 0.068^{+0.016}_{-0.017} & 0.064^{+0.011}_{-0.013} \vspace*{1.5pt}\\
\enddata
\tablecomments{Col. (1): name of the dwarf galaxy. Cols. (2) and (3): residual light fraction measurements within the one and two effective radii \myedit{isophotes}, respectively.}
\end{deluxetable}

On the one hand, these measurements omit the outermost region of the galaxy images for two main reasons. First, the noise level of the images starts dominating at two effective radii, so any measurements at larger radii become unreliable due to their big uncertainties. Second, the observed disk features do not extend beyond two effective radii, lifting the necessity of computing a residual light fraction in the outer regions. 

On the other hand, these measurements also omit the innermost region of the galaxy images, as the pixels that lie just at the center are affected by the PSF and usually contain some artifacts resulting from the IRAF \texttt{ellipse} fitting procedure. Thus, the central region is masked out by a circle of radius equal to 1.5 times the PSF FWHM, equivalent to $1.7-2.1$ arcsec in size depending on the PSF of the image. We clarify that this constitutes a negligible fraction of the effective radii, so no important information is being masked out; even when disk substructures are present at small radii. Nonetheless, we also note that central overdensities, such as nuclei, become masked by this procedure. \myedit{According to the morphological classification of \citet{Binggeli:1985}, the majority of our objects are nucleated, and only one object is non-nucleated (VCC1695). We agree with this classification based on our own visual inspection of galaxy unsharp masks that were created by using a very small smoothing kernel, in order to highlight small substructures such as nuclei.}

\begin{figure}[t]
\includegraphics[width=\columnwidth]{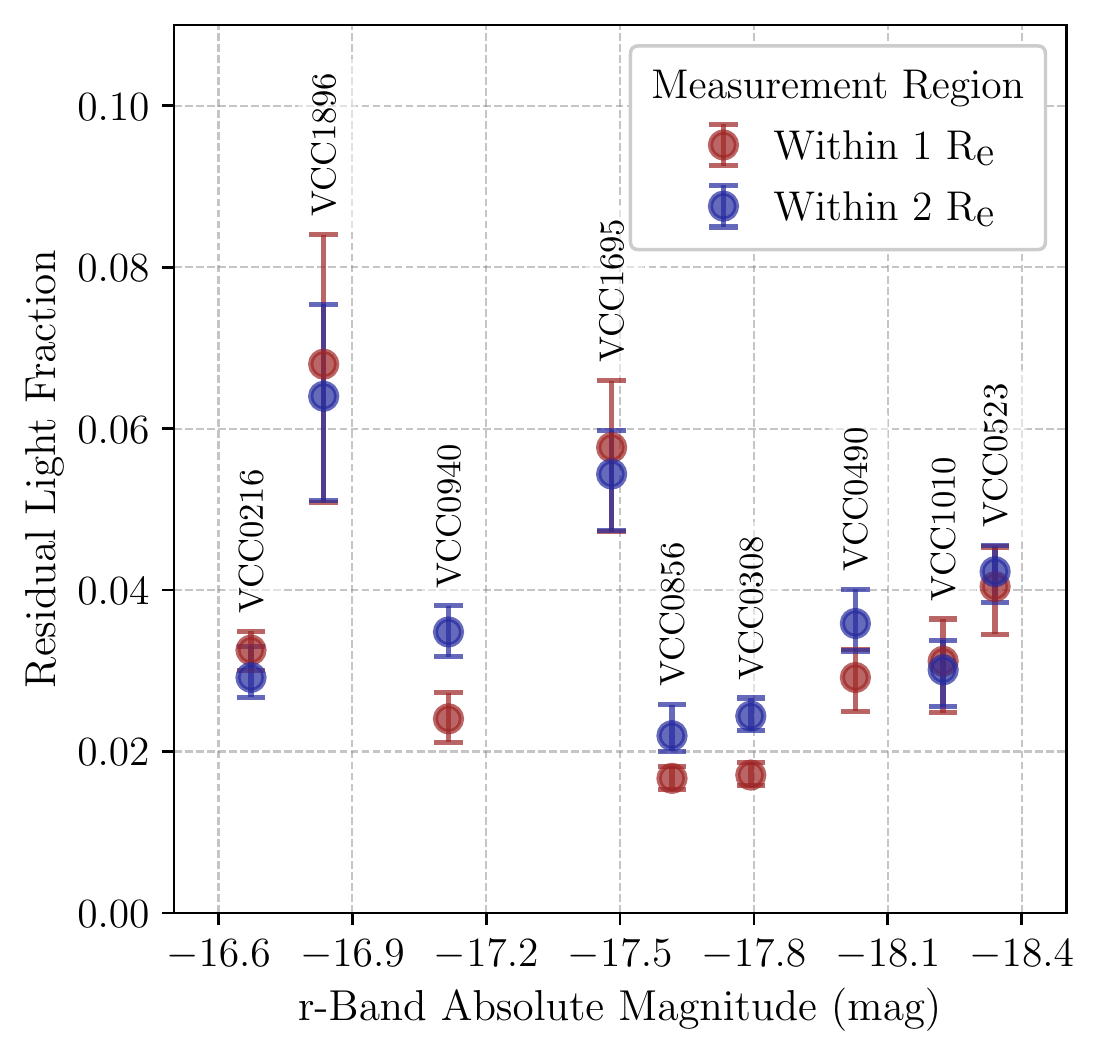}
\caption{Residual light fractions of the dwarf ETG sample, as a function of the total absolute magnitude of the galaxy in the $r$-band. The measurements are performed within one and two effective radii of the galaxies, which are shown as red and blue points, respectively. The error bars are given by the 16th and 84th percentiles of the distributions. Each pair of data points is labeled with the name of the corresponding galaxy.\label{fig:results:plot_res}}
\end{figure}

Additionally, any foreground and background sources, such as interloping stars or galaxies, are also masked out and omitted from the measurement. Finally, as the parameter setup of the residual method \myedit{(see Section \ref{subsec:method:config})} provides us with several residual images for each galaxy, we compute the residual light fraction by taking the median value of all measurements and, as their distribution is not necessarily symmetric, derive their uncertainties from the 16th and 84th percentiles.

\begin{figure*}[ht]
\includegraphics[width=\textwidth]{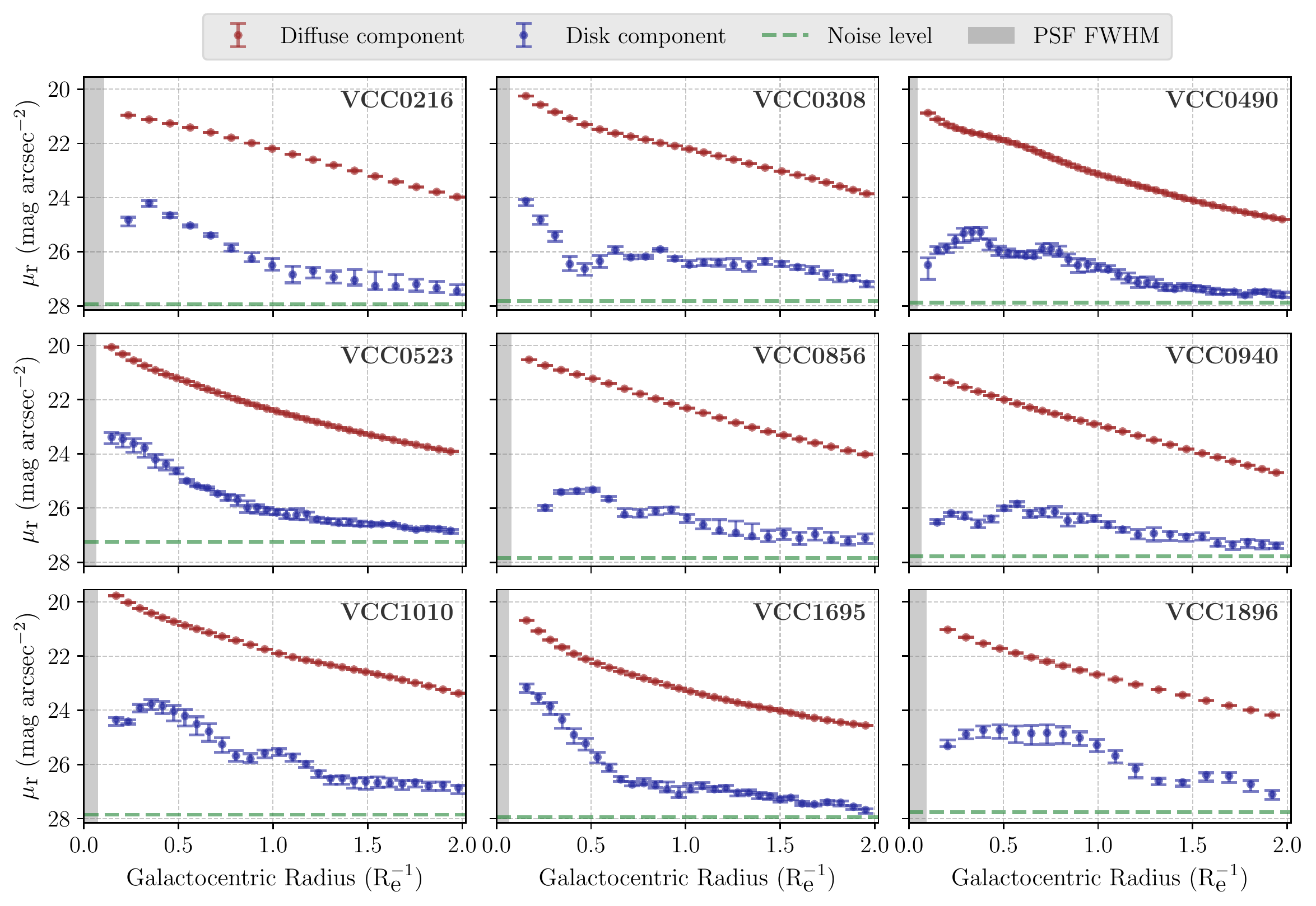}
\caption{Surface brightness profiles in the $r$-band of the diffuse and disk components of the dwarf ETG sample, shown as red and blue points, respectively. The measurements are performed in elliptical annuli of constant width that match the geometry of the isophotes of the diffuse component of the galaxy. The radial profiles are computed up to two effective radii of the galaxies. The central region is excluded due to the effects of the PSF, disregarding an amount equal to 1.5 times the PSF FWHM. The PSF FWHM is indicated by a gray shaded area. For reference, the background noise level is shown as a green dashed line.\label{fig:results:plot_SBprofiles}}
\end{figure*}

The residual light fraction measurements of the dwarf ETG sample are summarized in Table \ref{tab:results:res}. Overall, we find that throughout our sample the disk substructure contributes less than $7\%$ of the total galaxy light, supporting the fact that these features are indeed faint and lie hidden in a much brighter diffuse component. In particular, we find that the residual light fraction ranges between $1.7-6.8$\% within one effective radius, and between $2.2-6.4$\% within two effective radii. The differences between the one and two effective radii measurements for each galaxy are not significant, making either of them both good candidates to represent their residual light fraction. These results are visualized in Figure \ref{fig:results:plot_res}, which shows the residual light fraction as a function of the total $r$-band absolute magnitude of the galaxies. Even though the dwarf ETG sample spans a range of \myedit{almost} two magnitudes, we find no conclusive trend between the residual light fraction and the galaxy brightness. However, given that this is a small data set comprised of only nine dwarf galaxies, we are cautious to draw any conclusions.



As the diffuse and disk components of the galaxies have been separated into a model and a residual image, respectively, we can use these images to extract the surface brightness profile of each component. For this purpose, we construct elliptical annuli that match the ellipticity and position angle of the isophotes of the diffuse component of the galaxy, in order to reduce the effect of the disk features in driving the galaxy geometry. These elliptical annuli, which are concentric and of a constant 2 pixels ($1.42$ arcsec) in width, are then used to sample the galaxy model and galaxy residual images up to two effective radii. As a result, Figure \ref{fig:results:plot_SBprofiles} shows the surface brightness profiles of the diffuse and disk components of the dwarf ETG sample, which present some clear differences. On the one hand, the surface brightness profile of the diffuse component has a smooth appearance, \myedit{as its brightness declines steadily} with increasing galactocentric radius. For a more quantitative assessment, we run the 2D fitting algorithm GALFIT \citep{Peng:2002} on the galaxy model images. We thus find that the diffuse components are well described by a \myedit{single} S\'ersic profile \citep{Sersic:1968} of S\'ersic index $n$ ranging between $1.15-2.02$, with a median value of $n=1.43$. Therefore, the diffuse component of our dwarf ETGs is more consistent with having an exponential disk profile $(n=1)$ than a de Vaucouleurs profile \citep[$n=4$;][]{DeVaucouleurs:1948}, the latter being typical for giant elliptical galaxies. On the other hand, the profile of the disk component shows strong fluctuations that can be attributed to the \myedit{irregularity of the underlying substructures}, and its brightness tends to decline \myedit{less steeply} with galactocentric radius. \myedit{While we do not attempt to fit the residual images with GALFIT, given the numerous irregularities in their surface brightness profile, we note that there are some cases that would require at least two components to properly describe it (see, e.g., VCC0308 and VCC1695).} Furthermore, the surface brightness profile of the diffuse component is on average between three to four magnitudes brighter compared to the profile of the disk component at all radii. But, even though the disk component is comparatively very faint, its brightness within two effective radii of the galaxies is still above the background noise level.


Together, these results help build a picture in which the diffuse light and disk substructure light constitute two distinct galaxy components, as supported by their substantial differences in both brightness and appearance. However, the similarity of the residual light fraction measurements within one and two effective radii (see Table \ref{tab:results:res} and Figure \ref{fig:results:plot_res}) \myedit{could imply that the total physical extension of the diffuse component may not be too different from the one of the disk component.} In other words, these galaxies do not have inner or nuclear disks, but instead their disks are extended all along the galaxy body \myedit{(at least up to two effective radii, as can be appreciated in Figure \ref{fig:results:plot_SBprofiles})}. Having deep images, such as the ones at our disposal, was an essential requirement in order to reveal this behavior. Deep imaging allowed us to perform robust measurements towards the galaxy outskirts, which would not have been feasible with shallower imaging data.

\section{Testing the Method}\label{sec:rectests}

In order to assess both the reliability and relevance of the residual method, we construct a series of diagnostic tests. First, to evaluate its reliability, we test the accuracy with which it is able to recover the light contained in the disk features of a galaxy. Second, to evaluate its relevance, we compare its performance to the results obtained when other possible, alternative approaches are used to quantify the disk substructure light. We will show how these alternative approaches are not fundamentally designed to quantify substructures, thus justifying the creation of the residual method to fulfill that purpose. The aforementioned tests are carried out on a sample of mock galaxy images we build specifically for testing purposes. 

In Section \ref{subsec:rectests:setup}, we proceed to describe the creation of the mock galaxy images, followed by an explanation of the different methods to be tested. Then, in Section \ref{subsec:rectests:results}, we present the results obtained when applying these methods to the mock galaxy sample.

\subsection{Mock Galaxies and Tested Methods}\label{subsec:rectests:setup}

We first address how the mock galaxy sample is constructed. The mock galaxy images have an observational and not a simulated origin, in the sense that they are based on actual images from our dwarf ETG sample. In particular, we choose two dwarf galaxies each of which features one of the main signatures of a disk component: VCC0308, which has clear, wound-up spiral arms and no bar; and VCC1896, which has a bright, prominent bar and two faint, short spiral arms. To construct mock galaxy images from these two galaxies, we take the galaxy model and residual images obtained through the application of the residual method. Next, for each model and residual image pair, we measure the amount of light contained within their total galaxy area (defined as the region encompassed by the last valid isophote of the original galaxy image; see Section \ref{sec:data}). Then, we scale each image pair and add them in different relative fractions while keeping the total amount of light constant. As a result, we obtain a series of mock galaxy images \myedit{that share the same luminosity, but where some have fainter and some have brighter disk features. In particular, we choose} residual-to-total light fractions that encompass the range of empirical values obtained for the dwarf ETG sample (see Table \ref{tab:results:res}). Thus, for each of the two selected dwarf galaxies, we construct four mock galaxies by introducing residual light fractions equal to 2.5, 5.0, 7.5, and 10\% of the total light. The mock galaxy sample then consists of a total of eight mock galaxies. 

By having control over the introduced amount of residual light, we know the exact amount we should expect to recover when subjecting the mock galaxy images to a particular method. As part of the diagnostic test on the recovery efficiency of the residual light fraction, there are five different methods we want to compare, which are described below.\\


\noindent \textbf{1. Residual method.} The objective of the residual method, developed in this work, is to isolate the disk component of a galaxy in order to quantify the amount of light it contains. By applying it to the mock galaxy sample, we want to assess the accuracy with which the method is able to recover the introduced amount of residual light. For this test, we apply the complete procedure as described in Section \ref{subsec:method:steps}, and adopt the same parameter configuration we used for the dwarf ETG sample, described in Section \ref{subsec:method:config}.\\

\noindent \textbf{2. Residual method, no iterations.} The residual method includes an iterative procedure that allows the gradual separation of the diffuse and disk components of a galaxy into two separate images. To evaluate how important this iterative aspect is to the success of the method, we apply an incomplete, simplified version of the residual method to the mock galaxy images, in which there are no iterations. Therefore, we follow only Steps 1 through 5 described in Section \ref{subsec:method:steps}, and keep the first residual image that we obtain, thus omitting the iterative loop of the method. We once again adopt the same parameter configuration we used for the dwarf ETG sample, described in Section \ref{subsec:method:config}.\\

\noindent \textbf{3. Unsharp mask method.} The unsharp masking technique aims to bring forward any faint, underlying substructure that could be hiding in the bright, diffuse light of a galaxy image. An unsharp mask image is obtained by smoothing out the galaxy image, and then either dividing or subtracting the smoothed-out image from the original image. Both types of unsharp masking provide comparable results. As the process of smoothing spatially redistributes the light of the galaxy, thus altering the true radial light profile, this technique is only intended for visualization purposes, and not for quantification. However, in order to compare it to the residual method, we test the performance of unsharp masking as a means of quantifying disk substructure. First, using the IRAF \texttt{gauss} task, the mock galaxy images are smoothed out by adopting the same parameter configuration employed in Step 1 of the residual method (see Sections \ref{subsec:method:steps} and \ref{subsec:method:config}). Then, to create the unsharp mask images, the IRAF \texttt{imarith} task is used to subtract the smoothed-out images from the original images. We note that subtraction and not division is used to create these unsharp mask images, as through subtraction we obtain a measurable flux, while division provides a flux ratio. Finally, the resulting unsharp mask images are then treated as residual images, and are used to quantify the residual light.\\

\noindent \textbf{4. Fixed fit method.} When fitting a galaxy image, the most common objective is to model the brightness and geometry of the radial light profile as accurately as possible. In the ideal case that a model image is the perfect representation of a galaxy, by subtracting the model from the original image we obtain a residual image filled with zero values. Thus, the intended purpose of such a residual image is to assess the goodness of the fit, and not to quantify the disk substructure of the galaxy. However, we want to test what happens when the geometry of the galaxy is kept fixed during the fit \citep[as in][]{Lisker:2006a}, meaning that a single value for the ellipticity and position angle is imposed at all galactocentric radii. \myedit{As this geometrically simple fit does not take into account any radial twists and turns of the light profile, the complex geometry of the disk substructure will not be competently captured, so at least part of their light will appear in the resulting residual image.} To evaluate how this setup compares to the residual method, the mock galaxy images are fit using the IRAF \texttt{ellipse} task in fixed mode, by imposing at all galactocentric radii the geometry of the galaxy isophote at two effective radii. We adopt the same parameter configuration \myedit{and clipping criteria} used for sampling in the residual method, described in Section \ref{subsec:method:config}. Then, we construct the galaxy model images using the IRAF \texttt{bmodel} task, and subtract the model from the original images with the IRAF \texttt{imarith} task to create the galaxy residual images.\\

\noindent \textbf{5. Free fit method.} This approach follows what was described in the previous method, with one difference. \myedit{In this case, we perform an isophotal fit that allows the ellipticity and position angle to change freely at all galactocentric radii.} Therefore, the purpose of this method is to model the light profile of the galaxy as accurately as possible, \myedit{so the resulting residual image should contain little to no remaining substructure light}. To test how accurate the free fitting procedure is and how it compares to the residual method, we quantify any light that is left in the residual images of the mock galaxies. In this case, the mock galaxy images are fit using the IRAF \texttt{ellipse} task in free mode, by allowing the geometry of the galaxy isophotes to freely change radially. \myedit{We adopt the same sampling configuration described in the fixed fit method, and follow the same procedure to create the residual images.}

\subsection{Test Results}\label{subsec:rectests:results}

We now present the results obtained when applying these five different methods to the mock galaxy sample. We note that, as during the creation of the mock galaxy images the scaling and addition of the galaxy components are done in terms of the total galaxy light, all reported measurements are also performed in the total encompassing area of the galaxies.

\begin{figure}[t]
\includegraphics[width=\columnwidth]{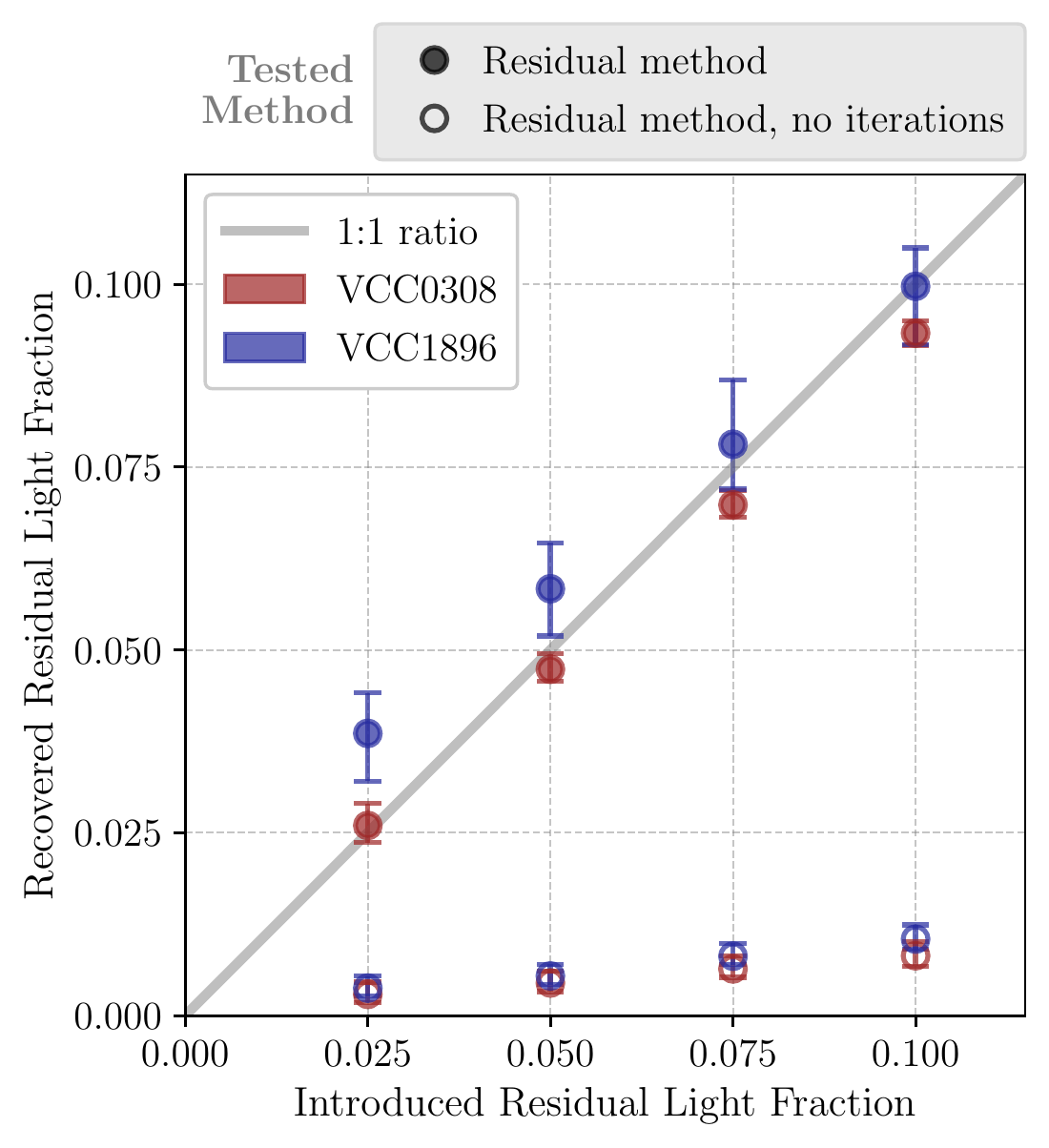}
\caption{Recovered vs. introduced residual light fractions obtained when applying the residual method to the mock galaxy images of VCC0308 (red points) and VCC1896 (blue points). The results of applying the complete residual method are shown by filled circles, while the results of applying an incomplete, simplified version of the residual method, in which there is no iterative procedure, are shown by empty circles. All measurements are performed within the total galaxy area, and their error bars are given by the 16th and 84th percentiles of the distribution. The one-to-one expected relation is shown as a gray line, representing the ideal case in which the introduced residual light fraction is recovered in its entirety.\label{fig:rectests:plot_res}}
\end{figure}

First, we assess the reliability of the residual method in recovering the introduced amount of disk substructure light. The recovered versus introduced residual light fractions are reported in Table \ref{tab:rectests:res}. The recovery efficiency of the residual method proves to be quite accurate, staying mostly within the uncertainty ranges. These results can be visualized in Figure \ref{fig:rectests:plot_res}, in which we notice an overall trend for the recovered fraction to slightly decrease with respect to its introduced fraction as the residual light increases (i.e., the filled circle data points start to go below the 1:1 expected relation). Therefore, it is reasonable to assume that the residual method only stays accurate when dealing with low levels of residual light \textendash as it was developed for that purpose\textendash, so its application is not advisable in cases with bright disk substructure (e.g., giant spiral galaxies). Additionally, the measurements present error bars of different sizes, which can be attributed to the morphology of the substructures that dominate in each mock galaxy. The galaxy VCC0308, with strong spiral arms, shows smaller uncertainties; while VCC1896, with a strong bar, presents instead bigger uncertainties. This may be an indication that the residual method is better suited at capturing the light contained in spiral arms than in prominent bar substructures.


\begin{deluxetable}{cCC}[t]
\tablecaption{Residual light fractions from applying the residual method to mock galaxies.\label{tab:rectests:res}}
\tablehead{
\colhead{Dwarf Galaxy} & \multicolumn2c{Residual Light Fraction} \vspace*{1pt}\\
\cline{2-3}
 & \colhead{Introduced} & \colhead{Recovered} \vspace*{-1pt}\\
\colhead{(1)} & \colhead{(2)} & \colhead{(3)} \vspace*{0.5pt}}
\startdata
VCC0308 & 0.025 & 0.026^{+0.003}_{-0.002} \\
 & 0.050 & 0.047^{+0.002}_{-0.002} \\
 & 0.075 & 0.070^{+0.002}_{-0.002} \\
 & 0.100 & 0.093^{+0.002}_{-0.002} \vspace*{1.5pt}\\
\cline{1-3}
VCC1896 & 0.025 & 0.039^{+0.006}_{-0.007} \\
 & 0.050 & 0.058^{+0.006}_{-0.006} \\
 & 0.075 & 0.078^{+0.009}_{-0.006} \\
 & 0.100 & 0.100^{+0.005}_{-0.008} \vspace*{1.5pt}\\
\enddata
\tablecomments{Col. (1): name of the dwarf galaxy used to construct the mock galaxy images. Col. (2): residual light fraction introduced into the mock galaxy. Col (3): residual light fraction recovered when applying the residual method.}
\end{deluxetable}

\begin{figure}[t]
\includegraphics[width=\columnwidth]{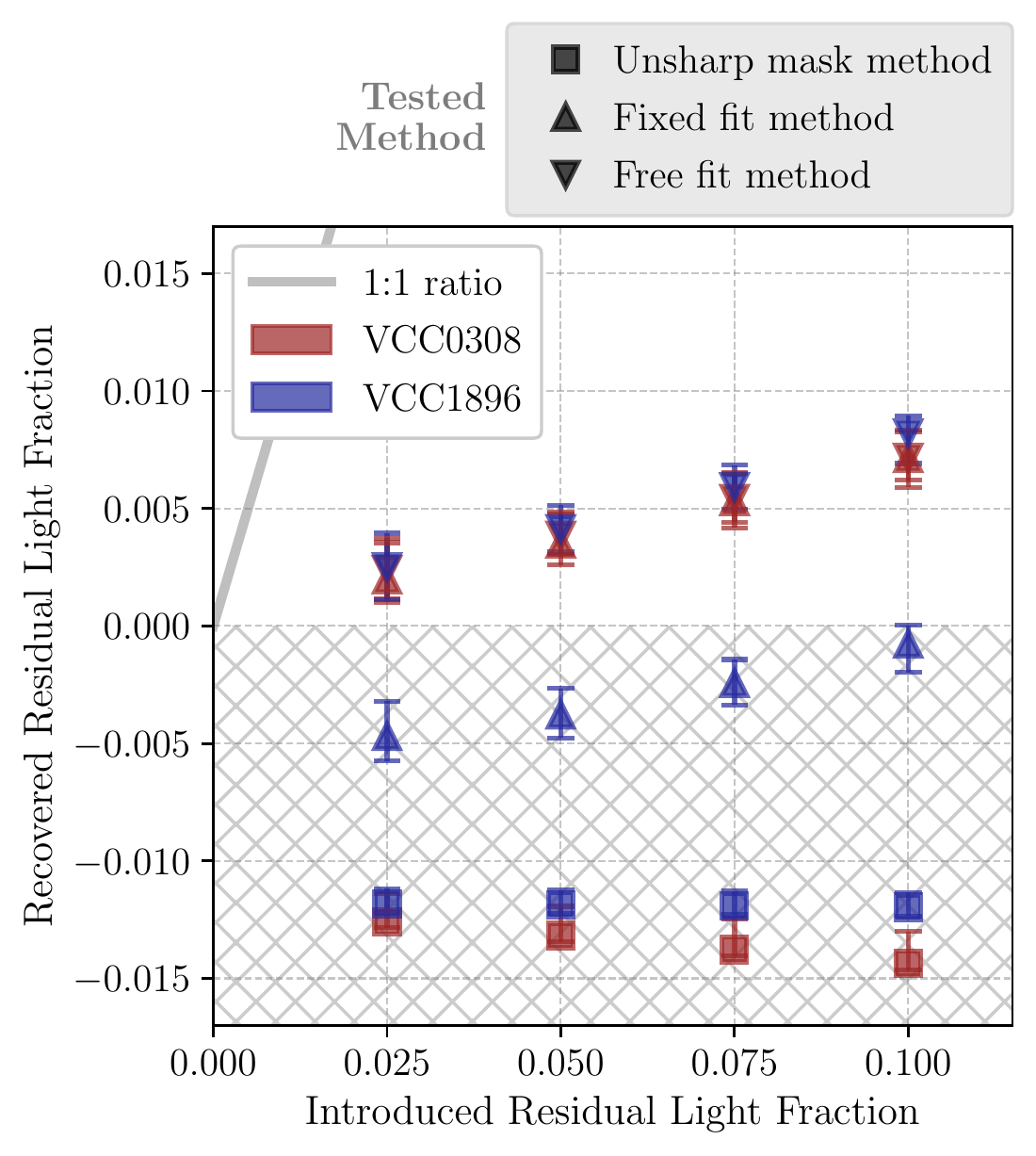}
\caption{Recovered vs. introduced residual light fractions obtained when applying alternative methods to the mock galaxy images of VCC0308 (red points) and VCC1896 (blue points). Different symbols represent the results of applying each method: the unsharp mask method (squares), the fixed fit method (up triangles), and the free fit method (down triangles). All measurements are performed within the total galaxy area, and their error bars are given by the 16th and 84th percentiles of the distribution. The hatched area delimits the negative recovery region of the plot, where the tested methods fail to recover a positive amount of residual light. The one-to-one expected relation is shown as a gray line, representing the ideal case in which the introduced residual light fraction is recovered in its entirety.\label{fig:rectests:plot_others}}
\end{figure}

In Figure \ref{fig:rectests:plot_res} we also compare the performance of the residual method with and without its iterative procedure. This test shows that the iterative aspect is fundamental: without iterating, the residual method is unable to properly isolate the light of the disk component from the diffuse component, resulting in a residual light fraction that is too low. \myedit{In other words, the galaxy residual image only contains a small fraction of the disk substructure light, while the majority of it is still blended with the diffuse light in the galaxy model image. Therefore, the complete, iterative version of the residual method is needed in order to provide accurate, reliable measurements of the faint disk substructure present in galaxies.}


\begin{figure*}[ht]
\includegraphics[width=\textwidth]{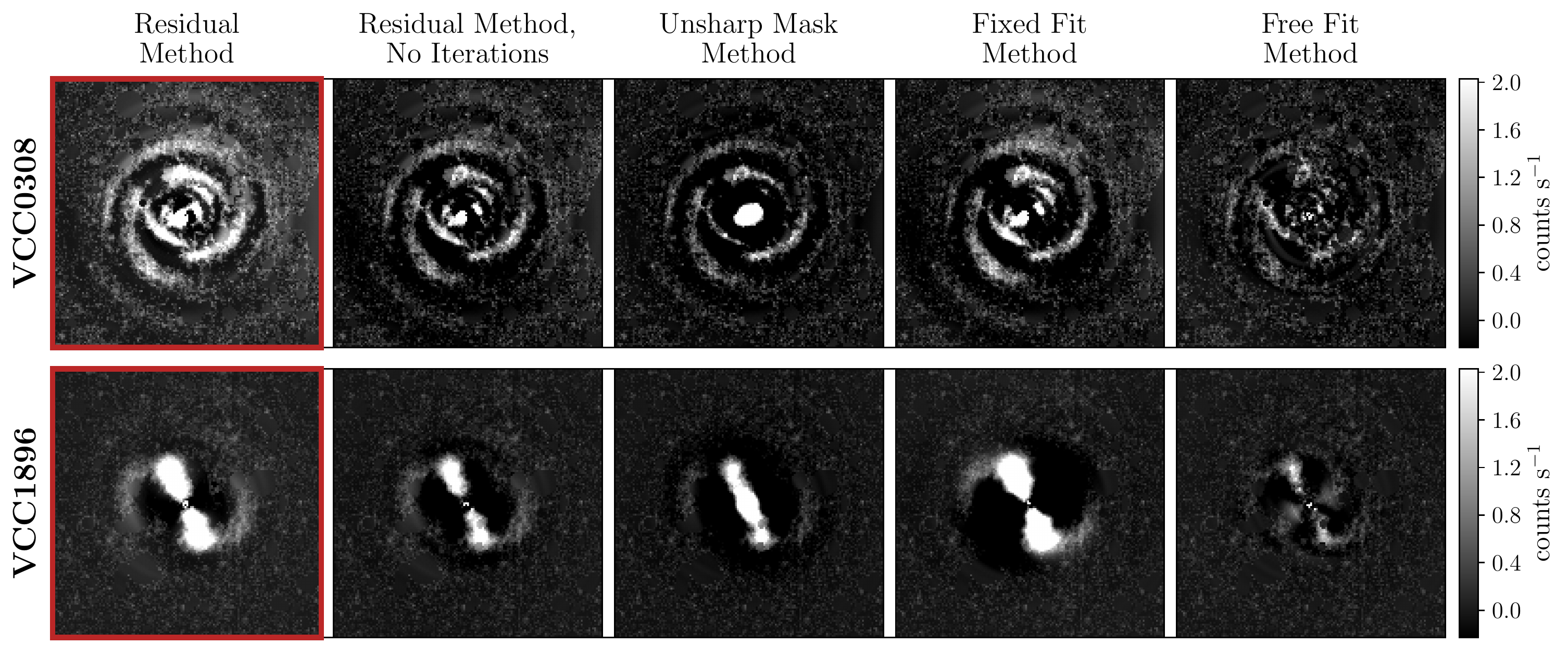}
\caption{Galaxy residual images obtained when applying different methods to the mock galaxy images based on VCC0308 (top row) and VCC1896 (bottom row). From left to right, the columns represent the different methods that have been tested: the residual method (highlighted in a red frame), the residual method with no iterations, the unsharp mask method, the fixed fit method, and the free fit method. The galaxy residual images that are shown correspond to the mock galaxies in which a residual-to-total light fraction of 10\% has been introduced. All images share the same gray-scale bar in units of counts s$^{-1}$, and are $100\times100$ arcsec$^{2}$ in size. North is up, east is to the left.\label{fig:rectests:imgs_res}}
\end{figure*}

Next, we evaluate the performance of alternative methods, normally not used for quantification purposes. In Figure \ref{fig:rectests:plot_others}, we present the recovered versus introduced residual light fractions when applying the unsharp mask method, the fixed fit method, and the free fit method to the mock galaxy sample. The main result is that these methods are unable to successfully recover the introduced amount of disk substructure light. First, for the unsharp mask method, we find that it fails to recover a positive amount of residual light. It is also mostly insensitive to the introduced residual light fraction, presenting a relatively flat relation. Then, for the fixed fit method, we find that its performance differs depending on the dwarf galaxy the mock images are based on. The mock galaxy images based on VCC0308 are in the positive recovery region of the plot, albeit with a very low recovery efficiency, while the mock galaxy images based on VCC1896 are in the negative recovery region of the plot. This discrepancy is a reflection of \myedit{how the complexity level of the galaxy geometry affects the goodness of the fit}. Finally, for the free fit method, we find that despite staying in the positive recovery region of the plot, it still does not recover enough light from the one introduced. In conclusion, these three methods cannot be used for quantification, and should instead be used for their intended purposes; namely, visualization \myedit{of faint embedded features} in the case of the unsharp mask method, and modeling \myedit{of the light distribution} in the case of the fixed and free fit methods. 


In order to visualize the results of these diagnostic tests, the galaxy residual images that are obtained by applying these five methods to the mock galaxy sample are displayed in Figure \ref{fig:rectests:imgs_res}. For illustration purposes, we only show the residual images of the mock galaxies in which a residual-to-total light fraction of 10\% has been introduced. Although at first sight the residual images of each mock galaxy may appear quite similar in general morphology, they present particularities that strongly differentiate one method from the next.


First, for the residual method, we observe that the disk features appear bright and with the correct geometry in the residual images, and that the negative values in the inter-arm galaxy regions have been minimized. In contrast, for the residual method with no iterations, the disk features present the same geometry but appear much fainter, while the inter-arm regions are dominated by negative values. \myedit{As previously stated, this is because the residual image is suffering from over-subtraction of the model image, which can only be alleviated by the iterative aspect of the method.}

We will now address the alternative methods. For the unsharp mask method, we observe that the disk substructure has the correct geometry in the residual images, albeit being too faint, while the inter-arm regions are dominated by negative values. Additionally, the bulge region appears artificially bright and prominent, which is an unwanted effect of the smoothing procedure: when smoothing, light from the brighter central region is redistributed towards the fainter galaxy outskirts. Thus, when subtracting this flattened light profile from the original galaxy image, the resulting residual image is unavoidably brighter in the central region and fainter in the outskirts. Therefore, when using unsharp masking for visualization purposes, this \myedit{artificial} effect should be kept in mind.

For the fixed fit method, we obtain different results depending on the complexity level of the galaxy geometry. On the one hand, VCC0308 is geometrically simple, as it has a face-on appearance and can thus be characterized by circular isophotes. Consequently, despite modeling the radial light profile with a single value for the ellipticity and position angle, the disk substructure in the residual image has the correct geometry. On the other hand, VCC1896 is geometrically more complex, as the inner region is dominated by a bar with highly elliptical isophotes, while the outer region is dominated by spiral arms with more circular isophotes. Therefore, a single value for the ellipticity and position angle fails to properly characterize the galaxy as a whole, and thus artificially alters the appearance of the disk substructure in the residual image. We observe that the bar appears more flared out, while the spiral arms are truncated. Additionally, and regardless of the specific galaxy geometry, the residuals of both mock galaxies are dominated by negative values. As a fixed fit approach has a high risk of underestimating the complex geometry of a galaxy, the user should be wary when using it for modeling purposes.

Finally, for the free fit method, we observe that the residual images contain only a very small fraction of the residual light. The remaining disk substructure either appears very faint, or is not present at all. This makes sense, as the objective of the free fit method is to model the light distribution of a galaxy as accurately as possible by allowing the ellipticity and position angle \myedit{of the isophotes} to change freely. We note that the IRAF \texttt{ellipse} fitting procedure has issues in fitting the bar of VCC1896, as an artificial feature appears perpendicular to the bar in the central region. In essence, these residual images then show the inherent limitations of the particular fitting procedure that is being implemented \myedit{(in this case, of the IRAF \texttt{ellipse} task)}, and can play a useful role when assessing the quality of the modeling.

In conclusion, it is clear from these qualitative and quantitative assessments that the residual method constitutes the best option to both reliably detect and accurately quantify faint disk substructure that is embedded in the bright diffuse body of a galaxy.

\section{Fourier Analysis}\label{sec:fourier}

In order to showcase the potential of the residual method, we subject the dwarf ETG sample to a Fourier analysis. The Fourier analysis of galaxy images can be used to both identify and characterize the different structural components that constitute a galaxy \citep{Elmegreen:1989,Rix:1995}. In particular, its application can also be extended to the dwarf regime, and has already been carried out on some of the dwarf galaxies belonging to our sample \myedit{as an attempt to quantify their faint disk features} \citep{Jerjen:2000,Barazza:2002}. In this work, we want to show that it is possible to go one step further, and perform a Fourier decomposition of the galaxy residual images obtained through the residual method. By using the galaxy residual images instead of the galaxy images themselves, we can obtain a quantitative characterization of the disk substructure that has already been identified and extracted from the galaxies and isolated in the residuals.

In the following Section \ref{subsec:fourier:formalism}, we introduce the Fourier decomposition formalism, and describe how to identify and characterize bar and spiral substructures. Then, in Section \ref{subsec:fourier:results}, we present the results obtained when applying the Fourier decomposition to the galaxy residual images of our sample.

\subsection{Fourier Formalism}\label{subsec:fourier:formalism}

Formally, the Fourier approach consists in the azimuthal spectral analysis of the surface brightness distribution of a galaxy image. The surface brightness profile is decomposed into a Fourier series, which takes the form
\begin{equation}\label{eq:fourier:decomp}
\Sigma(r,\theta) = \Sigma_{0}(r) + \sum_{m} A_{m}(r) \cos\left(m\theta - \Phi_{m}(r)\right),
\end{equation}
where $\Sigma(r,\theta)$ is the surface brightness of the galaxy expressed in polar coordinates, with $r$ being the radial distance and $\theta$ being the azimuthal angle. The associated Fourier amplitude and phase are given by $A_{m}$ and $\Phi_{m}$, respectively, where $m$ is the Fourier mode. $\Sigma_{0}(r)$ stands for the azimuthally averaged surface brightness profile.

The advantage of this approach is that the presence of disk features, such as bars and spiral arms, can be ascertained based on the behavior of specific Fourier parameters. On the one hand, the signature of a bar is typically associated with a prominence of even components, in particular of the mode $m=2$. Thus, a bar can be identified based on a constant phase $\Phi_{2}(r)$, which stays fixed with radius within a region of certain extent. On the other hand, the presence of spiral arms manifests itself through a linearly varying phase $\Phi_{m}(r)$ for an $m$-armed spiral mode. That is, a two-armed spiral feature would present a linear variation of $\Phi_{2}(r)$ with radius over an extended region.

First, we address the identification and characterization of bar-like substructure. When searching for a bar, we decide that a bar is present if the phase $\Phi_{2}(r)$ remains constant within $\pm 5$ degrees around the median over a large enough region, which we denominate as the bar region \citep{Kraljic:2012}. This definition is tuned to exclude spiral-like features, which typically vary by a few tens of degrees over several arcseconds. As an additional requirement, a bar region must present a peak in the normalized radial amplitude of mode $m=2$; $A_{2}(r)/A_{0}(r)$. The bar region should also start at galactocentric radii between $2.1-5.0$ arcsec ($3-7$ pixels) and cover a radial extension of at least $3.6$ arcsec (5 pixels). The minimum extent of the bar region is set such that it encompasses the bars we have already identified in our sample. We avoid the innermost galaxy region ($<2.1$ arcsec or 3 pixels), as central asymmetries can cause large variations in $\Phi_{2}(r)$ even for visually identified barred systems. We also estimate that the bar length measurements have an associated uncertainty of $0.71$ arcsec (1 pixel), which corresponds to the resolution of the Fourier algorithm, as it operates in radial bins of one pixel in size.

When the aforementioned criteria are met, the galaxy is classified as having a bar. To characterize the identified bar substructure, we proceed to quantify the bar strength, based on the definition proposed by \citet{Aguerri:1998},
\begin{equation}\label{eq:fourier:strength_bar_formal}
S_{\text{bar}} = \dfrac{1}{r_{\text{bar}}} \int_{r_{\text{start}}}^{r_{\text{end}}} \dfrac{A_{2}(r)}{A_{0}(r)} \, \text{d}r,
\end{equation}
where $r_{\text{start}}$ and $r_{\text{end}}$ are the inner and outer radius of the bar region, respectively; and $r_{\text{bar}}$ is the overall bar length $(r_{\text{bar}}=r_{\text{end}}-r_{\text{start}})$. $A_{0}(r)$ and $A_{2}(r)$ correspond to the Fourier amplitudes of modes $m=0$ and 2, respectively. However, as in practice the computation of the bar strength is a discrete measurement \myedit{in pixel-sized steps}, we use the following definition,
\begin{equation}\label{eq:fourier:strength_bar}
S_{\text{bar}} = \dfrac{1}{r_{\text{bar}}} \sum_{r=r_{\text{start}}}^{r_{\text{end}}} \dfrac{A_{2}(r)}{A_{0}(r)}.
\end{equation}

This definition is implemented with a small adjustment: we always assign $A_{0}(r)$ as the amplitude of mode $m=0$ of the original galaxy image. \myedit{In other words, when analyzing either the original image or the residual image, $A_{2}(r)$ is always normalized by the $A_{0}(r)$ of the original image.} Additionally, to assess the influence of the innermost galaxy region \myedit{that is being excluded ($r<2.1$ arcsec)}, we also provide in each case an alternative bar strength measurement that starts instead at $r_{\text{start}}=0$.

We also want to quantify the disk substructure of the whole galaxy sample, independently of whether they are barred or non-barred systems. For this purpose, we also consider the overall contribution of their mode $m=2$ \myedit{relative to their mode $m=0$.} By adapting Equation \ref{eq:fourier:strength_bar}, we thus define the \myedit{total} strength of mode $m=2$ as
\begin{equation}\label{eq:fourier:strength_m2}
S_{2} = \dfrac{1}{R_{e}} \sum_{r=0}^{R_{e}} \dfrac{A_{2}(r)}{A_{0}(r)},
\end{equation}
where $R_{e}$ is the effective radius of the galaxy, as provided in Table \ref{tab:data:gal_params}. As before, \myedit{when either analyzing the original or the residual image,} $A_{0}(r)$ always corresponds to the amplitude of mode $m=0$ of the original image.

Next, we address the analysis of spiral-like substructure. From the visual inspection of the galaxy residual images, we observe only two-armed spiral features present across the sample. Thus, we search for regions in which the phase $\Phi_{2}(r)$ varies linearly with \myedit{galactocentric} radius, which we then denominate as spiral arm regions. \myedit{In order to characterize the identified spiral features, we proceed to compute their pitch angles.} At a given radius, the pitch angle $\psi(r)$ is defined as the angle between the tangent of a spiral arm and the tangent of a circle, and is a measure of the tightness of the winding of the spiral arm \citep{Binney:2008}. Tightly wound spiral arms have small pitch angles, while loosely wound spiral arms have big pitch angles. Formally, the pitch angle can vary between $-90<\psi(r)<90$ degrees, where the sign indicates the direction of rotation of the spiral arm; positive being clockwise winding, and negative being counterclockwise winding.

\myedit{As our sample is constituted by real galaxies} that do not have perfect logarithmic spiral features, their pitch angles are expected to vary with radius. Taking this into consideration, we linearly fit \myedit{in the mode $m=2$ phase plot each spiral arm segment separately}, in the form $\Phi_{2}(r) = ar + b$; where $a$ is the slope and $b$ the intercept. From this linear fit, we can then derive a pitch angle for \myedit{any given radius $r$ of the spiral}, given by
\begin{equation}\label{eq:fourier:pitch}
\psi = \arctan \left(\dfrac{\text{d}r}{r\,\text{d}\Phi_{2}}\right) = \arctan \left(\dfrac{1}{a r}\right),
\end{equation}
where $\text{d}\Phi_{2}/\text{d}r$ is the slope $a$ of the linear relation. \myedit{This way, we obtain a series of pitch angles (one for each radial step) for the spiral arm regions. Finally, by taking the median of all pitch angle values, we compute a representative pitch angle for each galaxy,} and derive their uncertainties from the 16th and 84th percentiles.

As a final remark, the Fourier analysis we carry out does not make any assumption on the possible inclination angles of the dwarf ETGs. Consequently, derived quantities such as the bar length and median spiral arm pitch angle correspond to projected measurements.

\subsection{Results of the Fourier Analysis}\label{subsec:fourier:results}

\begin{figure*}[p]
\begin{center}
\gridline{\fig{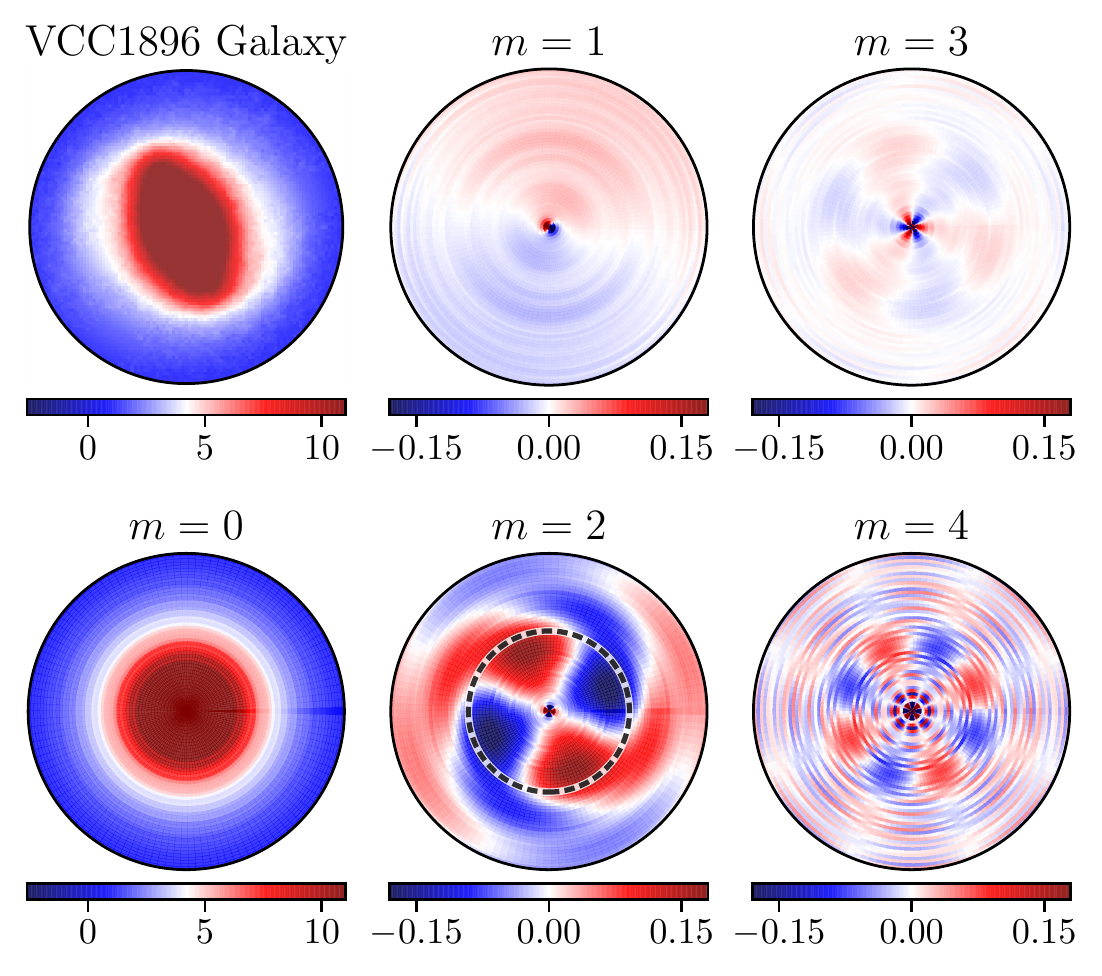}{0.46\textwidth}{}
          \fig{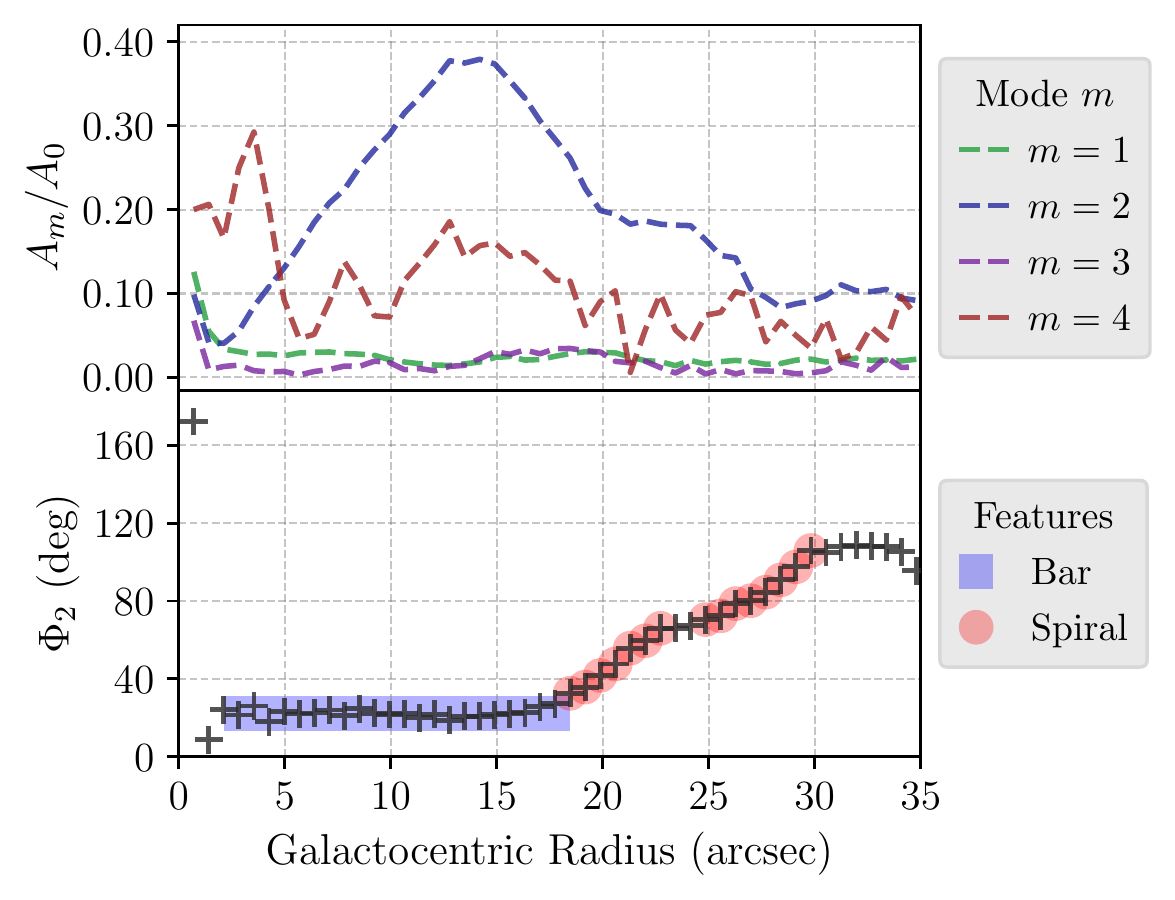}{0.53\textwidth}{}}
          \vspace*{-18pt}
\caption{{Fourier mode maps and Fourier parameters corresponding to the original galaxy image of VCC1896. \textit{Left circular panels}: the original galaxy image and maps of Fourier modes $m=0$ through $m=4$ are shown. \myedit{The Fourier maps of modes $m=1$ through $m=4$ have been normalized by the map of mode $m=0$. Each image has its own color bar; the original galaxy image and the map of mode $m=0$ are in units of counts s$^{-1}$, while the maps of modes $m=1$ through $m=4$ are unitless.} The dashed circle in the $m=2$ map marks the detected end radius of the bar. The images display a circular area of 35 arcsec in radius. North is up, east is to the left. \textit{Right rectangular panels}: the Fourier parameters, amplitude and phase, as a function of galactocentric radius are shown. On the top panel, \myedit{we show the amplitude of modes $m=1$ through $m=4$, which have been normalized by the amplitude of mode $m=0$.}} On the bottom panel, we show the phase of mode $m=2$, measured counterclockwise from the $+y$-axis. The detected bar-like substructure is highlighted by a blue rectangle, while the data points associated with spiral-like substructure are highlighted by red circles.\label{fig:fourier:vcc1896_gal}}
\end{center}
\end{figure*}

\begin{figure*}[ht]
\begin{center}
\gridline{\fig{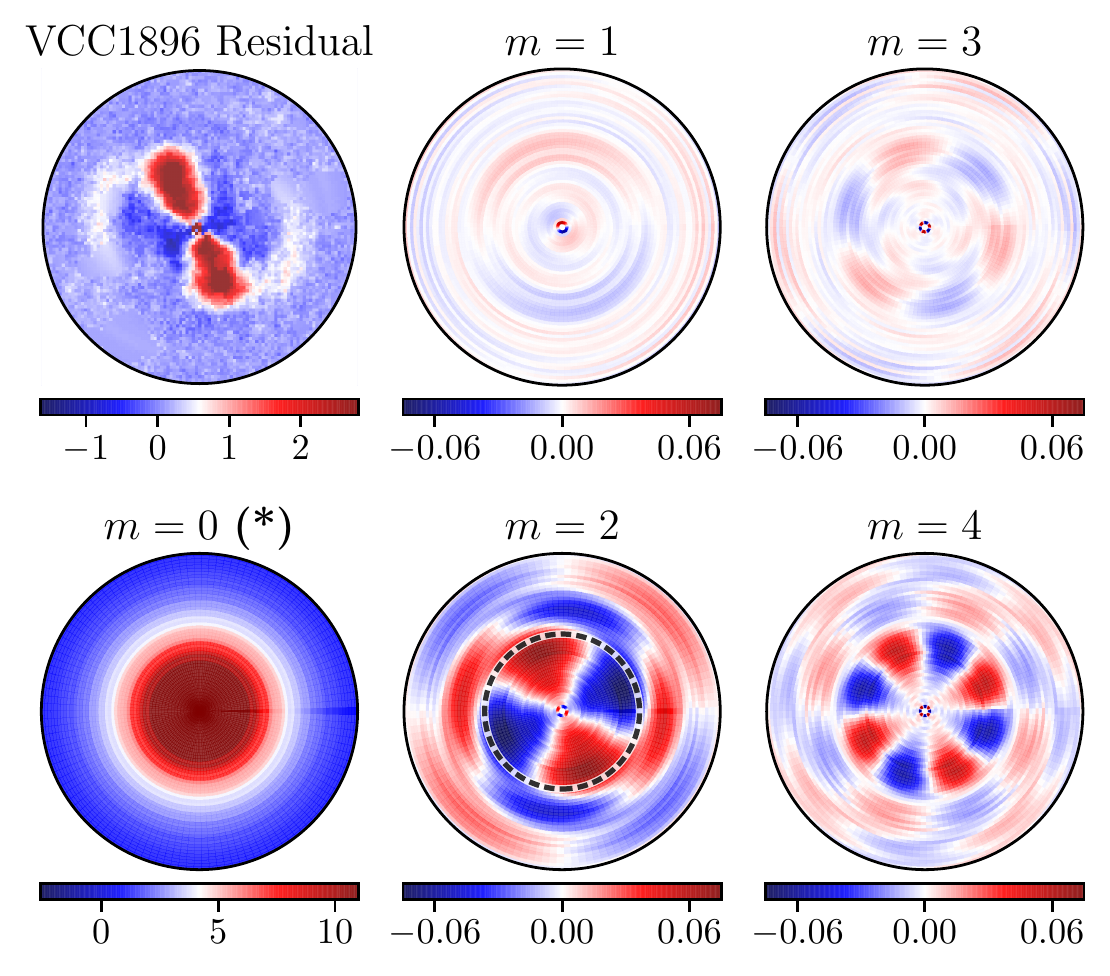}{0.46\textwidth}{}
          \fig{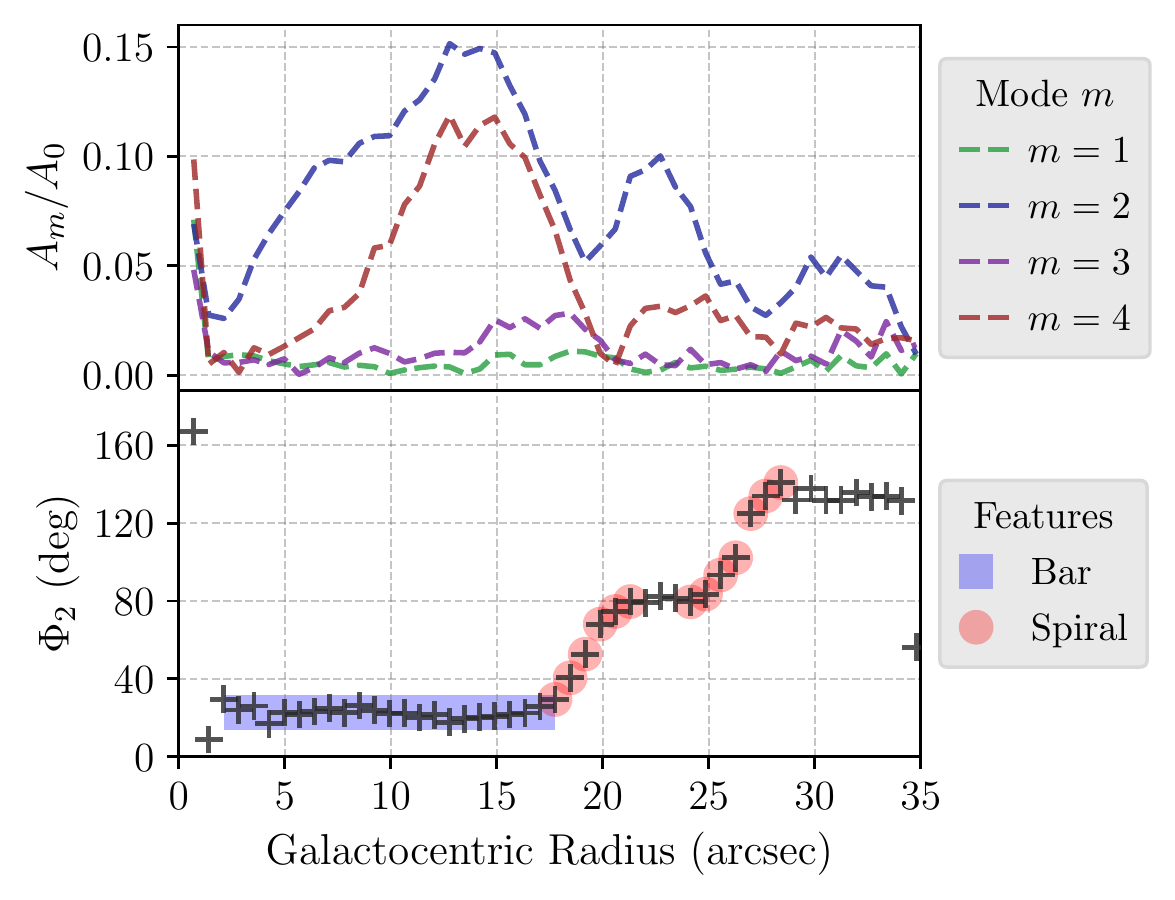}{0.53\textwidth}{}}
          \vspace*{-18pt}
\caption{{As in Figure \ref{fig:fourier:vcc1896_gal}, but for the galaxy residual image of VCC1896. \myedit{We note that the Fourier maps of modes $m=1$ through $m=4$ have been normalized by the map of mode $m=0$ of the original galaxy image, which is once again displayed here and marked with an asterisk (*). Likewise, the amplitude of modes $m=1$ through $m=4$ have been normalized by the amplitude of mode $m=0$ of the original galaxy image. The galaxy residual image is in units of counts s$^{-1}$. Both bar-like and spiral-like substructures are detected.}}\label{fig:fourier:vcc1896_res}}
\end{center}
\end{figure*}

\begin{figure*}[ht]
\begin{center}
\gridline{\fig{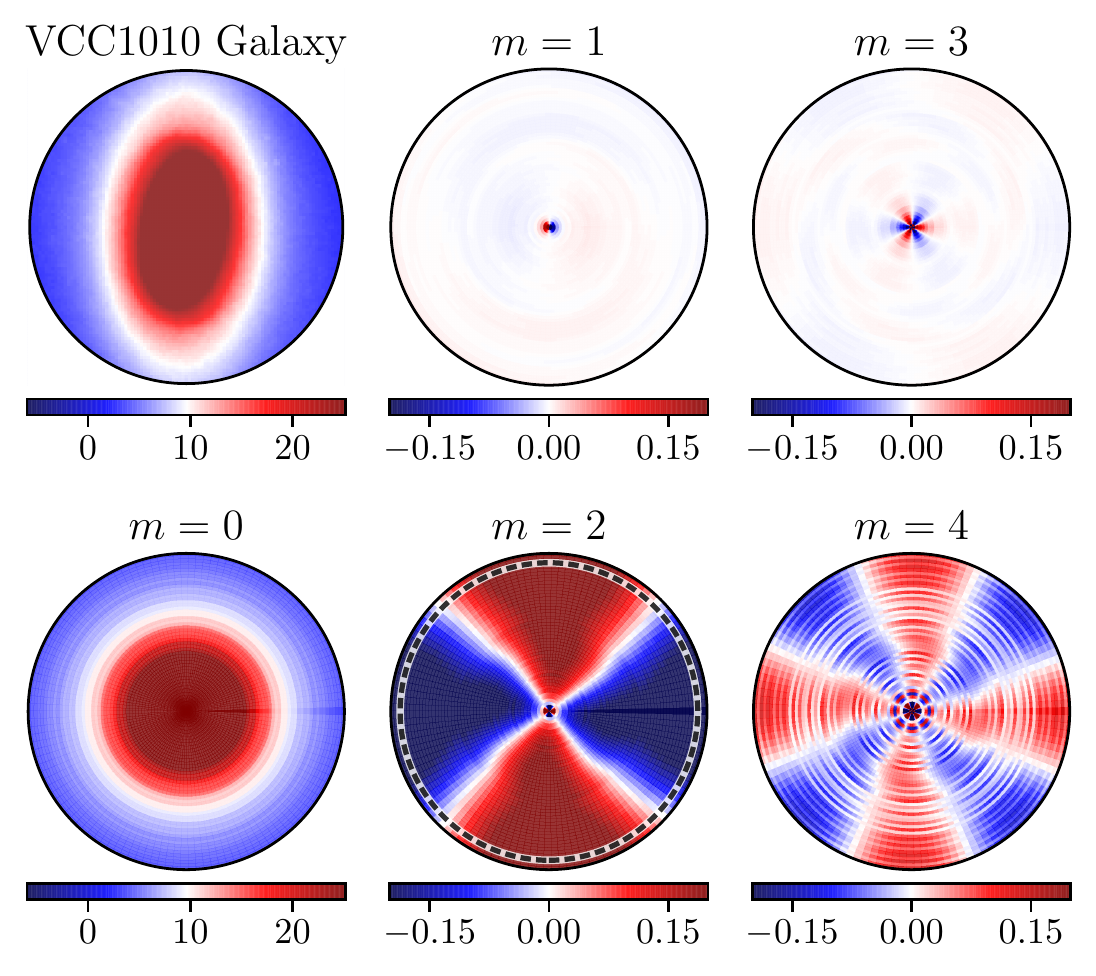}{0.46\textwidth}{}
          \fig{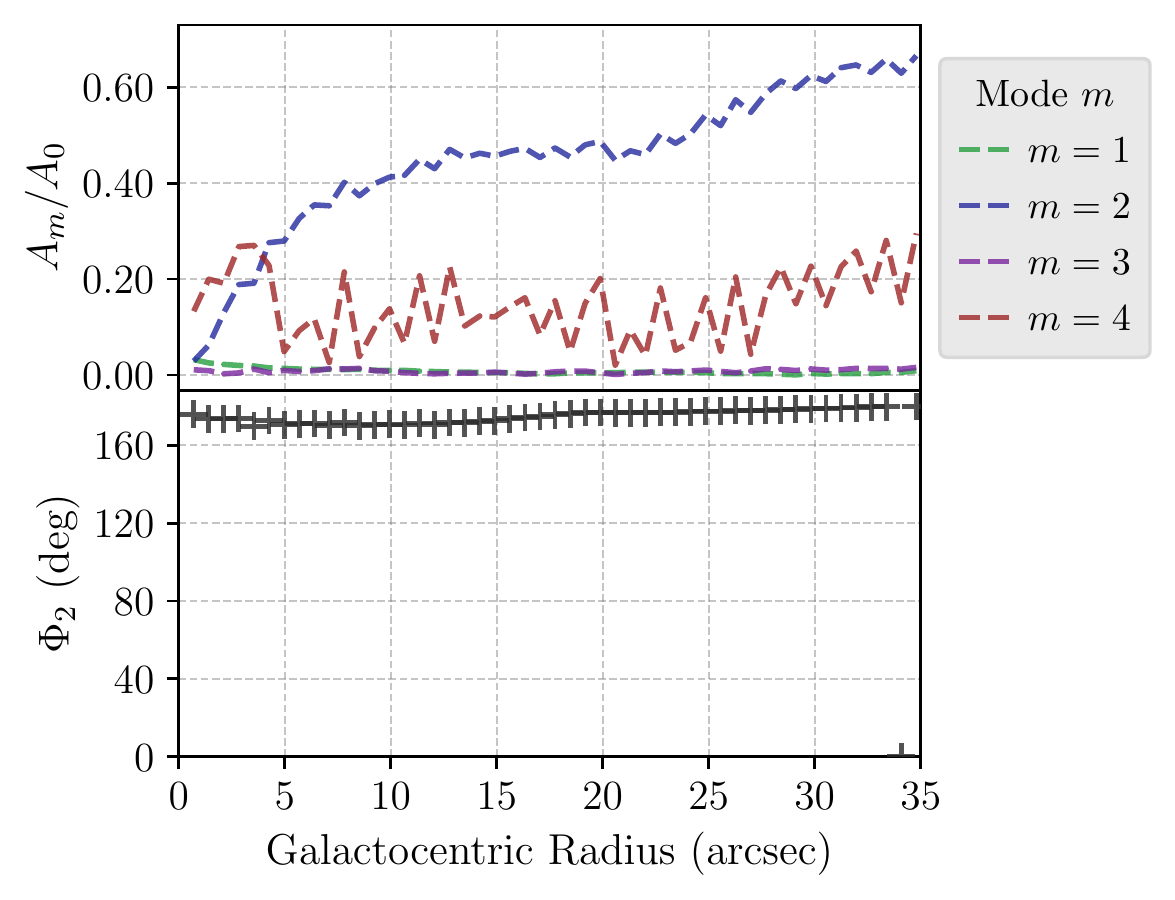}{0.53\textwidth}{}}
          \vspace*{-18pt}
\caption{{As in Figure \ref{fig:fourier:vcc1896_gal}, but for the original galaxy image of VCC1010. No bar-like nor spiral-like substructures are detected, while instead we identify a strong triaxial feature.}\label{fig:fourier:vcc1010_gal}}
\end{center}
\end{figure*}

\begin{figure*}[ht]
\begin{center}
\gridline{\fig{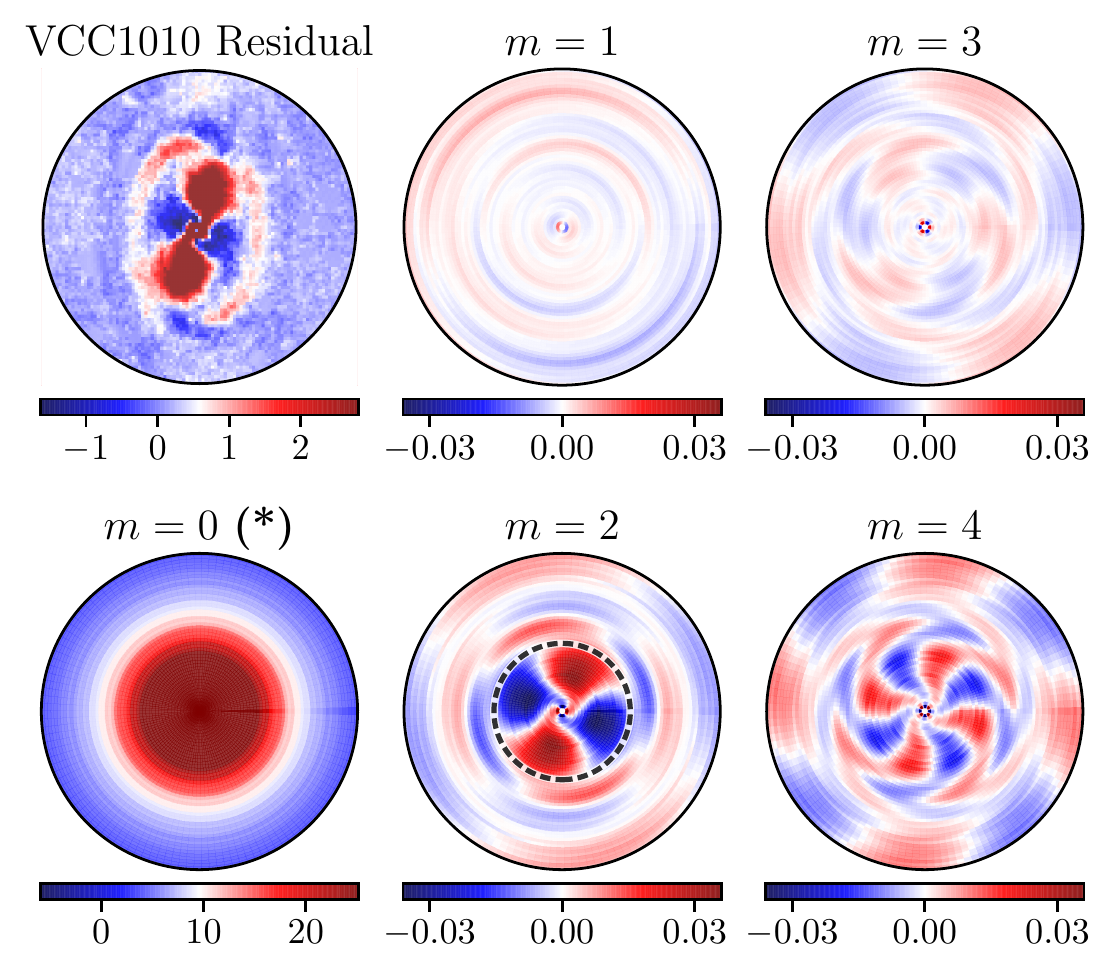}{0.46\textwidth}{}
          \fig{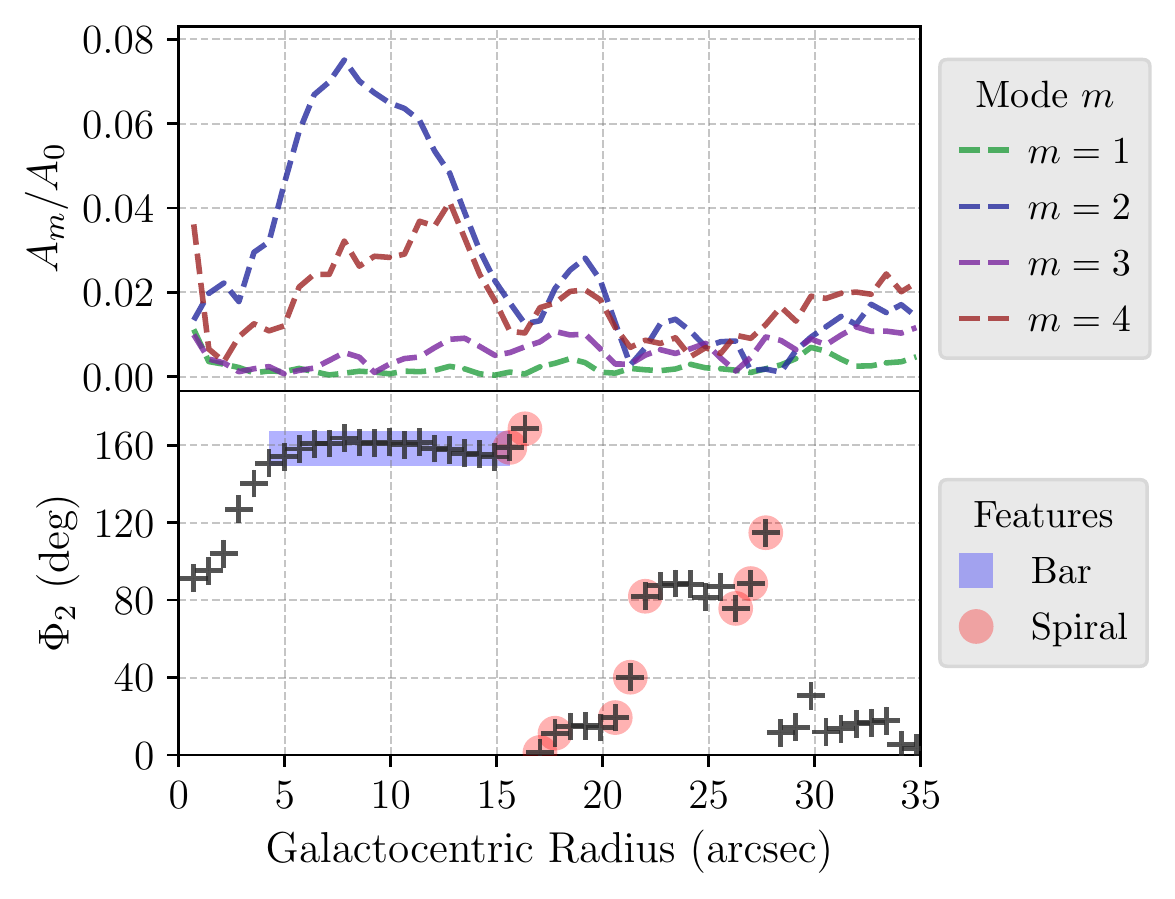}{0.53\textwidth}{}}
          \vspace*{-18pt}
\caption{{As in Figure \ref{fig:fourier:vcc1896_res}, but for the galaxy residual image of VCC1010. Both bar-like and spiral-like substructures are detected.}\label{fig:fourier:vcc1010_res}}
\end{center}
\end{figure*}

To prove how the aforementioned analysis of the bar and spiral substructure benefits from using the galaxy residual images instead of the original images themselves, we provide two examples: the galaxies VCC1896 and VCC1010. These two galaxies are chosen because they present both similarities and differences, which makes them ideal for a comparison. They have a similar morphology, with a strong bar feature dominating their disk substructure. However, they differ in their residual-to-total amount of light, with VCC1896 having a residual light fraction more than two times larger than VCC1010 (see Table \ref{tab:results:res}).

On the one hand, \myedit{the galaxy VCC1896 has comparatively brighter substructure, and the Fourier analysis} is successful when applied to both the original image (Figure \ref{fig:fourier:vcc1896_gal}) and the residual image (Figure \ref{fig:fourier:vcc1896_res}). In both cases, the even modes dominate the surface brightness distribution, and the disk substructure of the galaxy is clearly captured in the mode $m=2$ map. The amplitude plots also show a peak of the normalized $A_{2}$ amplitude, which is a signature of bar-like substructure, and the phase $\Phi_{2}$ plots allow an unambiguous identification of both the bar and spiral regions. \myedit{We would like to highlight that a bar of almost identical length is detected in both cases, and that the position of the normalized $A_{2}$ peak is preserved. This works as a confirmation of the robustness of the residual method, proving that it is able to accurately extract any disk substructure and isolate it in a residual image.}

On the other hand, \myedit{the galaxy VCC1010 has comparatively fainter substructure, and the Fourier analysis} is unsuccessful in extracting any disk features when applied to the original image (Figure \ref{fig:fourier:vcc1010_gal}), but it succeeds when applied to the residual image (Figure \ref{fig:fourier:vcc1010_res}). For the original image, we observe no peak in the normalized $A_{2}$ amplitude, so the region of constant $\Phi_{2}$ that we detect does not correspond to a bar, but instead to a strong triaxial feature. We are also unable to detect any spiral-like features. In contrast, in the case of the residual image the disk substructure has already been extracted and isolated, so this time the Fourier decomposition succeeds in describing it. We can clearly recognize the bar and spiral regions in the mode $m=2$ map, confirm their behavior in the amplitude plot, and constrain them based on the phase $\Phi_{2}$ plot. In conclusion, when the residual-to-total light is too low \myedit{as in the case of VCC1010}, the Fourier decomposition fails to extract the disk substructure that lies hidden in the much brighter main body of the galaxy. For this reason, the Fourier analysis that we subsequently carry out will only consider the residual images of the dwarf ETG sample.


The results obtained from the Fourier analysis of the galaxy residual images are presented in Table \ref{tab:fourier:params}. Overall, we find six galaxies that are barred (VCC0490, VCC0523, VCC0940, VCC1010, VCC1695, and VCC1896), and three galaxies that are non-barred (VCC0216, VCC0308, and VCC0856). For the barred systems, we characterize the bar substructure by deriving its extension \textendash start radius $(r_{\text{start}})$, end radius $(r_{\text{end}})$, and length $(r_{\text{bar}})$\textendash\ and orientation \textendash phase $(\Phi_{\text{bar}})$\textendash\ from the $\Phi_{2}(r)$ plot, and compute its strength $(S_{\text{bar}})$ through Equation \ref{eq:fourier:strength_bar}. Then, for the full sample, we compute the mode $m=2$ strength $(S_{2})$ through Equation \ref{eq:fourier:strength_m2}. We also derive the tightness \textendash median pitch angle $(\psi)$\textendash\ of their spiral substructure based on the $\Phi_{2}(r)$ plot and by applying Equation \ref{eq:fourier:pitch}.

The Fourier analysis shows that the identified bar features are between $5.7-15.6$ arcsec in projected angular length, with a median of 9.6 arcsec. Assuming a Virgo cluster distance of 16.52 Mpc, derived from a Virgo distance modulus of $31.09 \pm 0.15$ mag \citep{Blakeslee:2009}, these angular sizes are equivalent to a projected physical length ranging between $455-1251$ pc, with a median of 768 pc. We also find that the strength of the bar features ranges between $3.5-12.7$\%. The alternative measurements starting at $r_{\text{start}}=0$ only alter the result by a few percent.

For the identified spiral features, the absolute value of their median projected pitch angles ranges between $4.5-25.4$ degrees. We note that \citet{Lisker:2006a} reports pitch angles in agreement with ours for the four galaxies we have in common (VCC0308, VCC0490, VCC0856, and VCC1896; refer to their Figure 8), despite using a completely different approach. The majority of our sample presents small pitch angles, and thus tightly wound spiral arms, typically found in spiral galaxies of Hubble early-type \citep[S0$-$Sb;][]{Ma:1999}. However, we also have some galaxies with comparatively bigger pitch angles, being instead more consistent with spiral late-types (Sc$-$Sd). Therefore, even though the dwarf ETG sample tends towards small pitch angles, the broad range they cover could be a possible indication of their diverse origins or evolutionary histories.

Finally, the full sample presents a strength of the mode $m=2$ that lies in the range between $2.0-8.3$\%. We can directly compare these Fourier strengths with the residual strengths obtained through the residual method, previously reported in Table \ref{tab:results:res} and Figure \ref{fig:results:plot_res}. Within the same measurement region of one effective radius, the residual method finds residual-to-total light fractions that range between $1.7-6.8$\%. Consequently, the Fourier and residual methods reach similar results, even though their quantification techniques are fundamentally different. Together, they support a picture in which the underlying disk substructure light only constitutes a few percent of the much brighter and dominant diffuse light of these galaxies.

\begin{deluxetable*}{cCDDDCCDL@{\hspace{1em}}}[ht]
\tablecaption{Quantities from the Fourier component analysis of the galaxy residual images.\label{tab:fourier:params}}
\tablehead{
\colhead{Dwarf Galaxy} & \multicolumn5c{Bar Length} & \multicolumn2c{Bar Phase} & \colhead{Bar Strength} & \colhead{Mode 2 Strength} & \multicolumn2c{Pitch Angle} \vspace*{1pt}\\
\cline{2-6}
 & \text{Start} & \multicolumn2c{End} & \multicolumn2c{Length} & \multicolumn2c{} & & & \multicolumn2c{} \\
 & \colhead{(arcsec)} & \multicolumn2c{(arcsec)} & \multicolumn2c{(arcsec)} & \multicolumn2c{(deg)} & & & \multicolumn2c{(deg)} \\
\colhead{(1)} & \colhead{(2)} & \multicolumn2c{(3)} & \multicolumn2c{(4)} & \multicolumn2c{(5)} & \colhead{(6)} & \colhead{(7)} & \multicolumn2c{(8)} \vspace*{0.5pt}}
\decimals
\startdata
VCC0216 & \text{---} & \multicolumn2c{\text{---}} & \multicolumn2c{\text{---}} & \multicolumn2c{\text{---}} & \text{---} & 0.046 & \hspace*{4pt}23.1 & \hspace*{-32.5pt}^{+5.8}_{-4.0} \\
VCC0308 & \text{---} & \multicolumn2c{\text{---}} & \multicolumn2c{\text{---}} & \multicolumn2c{\text{---}} & \text{---} & 0.020 & 11.2 & \hspace*{-32.5pt}^{+9.0}_{-2.3} \\
VCC0490 & $7.10\pm0.71$ & 12.78\pm0.71 & 5.68\pm0.71 & 68.4 & 0.036\,(0.032) & 0.030 & 17.5 & \hspace*{-32.5pt}^{+21.2}_{-7.8} \\
VCC0523 & $2.13\pm0.71$ & 8.52\pm0.71 & 6.39\pm0.71 & 19.9 & 0.083\,(0.067) & 0.045 & 25.4 & \hspace*{-32.5pt}^{+15.8}_{-9.2} \\
VCC0856 & \text{---} & \multicolumn2c{\text{---}} & \multicolumn2c{\text{---}} & \multicolumn2c{\text{---}} & \text{---} & 0.022 & 12.8 & \hspace*{-32.5pt}^{+5.2}_{-5.2} \\
VCC0940 & $7.81\pm0.71$ & 22.72\pm0.71 & 14.91\pm0.71 & 20.1 & 0.035\,(0.028) & 0.026 & 13.0 & \hspace*{-32.5pt}^{+0.5}_{-0.4} \\
VCC1010 & $4.26\pm0.71$ & 15.62\pm0.71 & 11.36\pm0.71 & 158.4 & 0.052\,(0.043) & 0.043 & 4.5 & \hspace*{-32.5pt}^{+8.0}_{-0.9} \\
VCC1695 & $2.13\pm0.71$ & 9.94\pm0.71 & 7.81\pm0.71 & 22.1 & 0.127\,(0.109) & 0.068 & 15.6 & \hspace*{-32.5pt}^{+3.2}_{-6.0} \\
VCC1896 & $2.13\pm0.71$ & 17.75\pm0.71 & 15.62\pm0.71 & 22.8 & 0.103\,(0.095) & 0.083 & 11.0 & \hspace*{-32.5pt}^{+6.8}_{-1.2}\vspace*{1.5pt}\\
\enddata
\tablecomments{Col. (1): name of the dwarf galaxy. Cols. (2) to (4): projected bar length quantities; the bar starting radius $(r_{\text{start}})$, bar ending radius $(r_{\text{end}})$, and overall bar length $(r_{\text{bar}}=r_{\text{end}}-r_{\text{start}})$, respectively. Col. (5): bar phase $(\Phi_{\text{bar}})$, measured counterclockwise from the $+y$-axis (north towards east of the images). Col. (6): bar strength $(S_{\text{bar}})$, as defined in Equation \ref{eq:fourier:strength_bar}. In parenthesis, an alternative bar strength measurement starting instead at $r_{\text{start}}=0$. Col. (7): strength of mode $m=2$ $(S_{2})$, as defined in Equation \ref{eq:fourier:strength_m2}. Col. (8): absolute value of the median projected spiral arm pitch angle $(|\psi|)$, as defined in Equation \ref{eq:fourier:pitch}.}
\end{deluxetable*}


\begin{deluxetable}{cCC}[ht]
\tablecaption{Bar and spiral light fractions of the dwarf ETG sample.\label{tab:fourier:bar_sp_rlf}}
\tablehead{
\colhead{Dwarf Galaxy} & \multicolumn2c{Residual Light Fraction} \vspace*{1pt}\\
\cline{2-3}
 & \colhead{Bar} & \colhead{Spiral} \vspace*{-1pt}\\
\colhead{(1)} & \colhead{(2)} & \colhead{(3)} \vspace*{0.5pt}}
\startdata
VCC0216 & \text{---} & 0.029^{+0.004}_{-0.002} \\
VCC0308 & \text{---} & 0.024^{+0.002}_{-0.002} \\
VCC0490 & 0.028^{+0.005}_{-0.006} \left(0.022^{+0.004}_{-0.004}\right) & 0.039^{+0.006}_{-0.003} \\
VCC0523 & 0.050^{+0.010}_{-0.012} \left(0.050^{+0.010}_{-0.012}\right) & 0.039^{+0.003}_{-0.002} \\
VCC0856 & \text{---} & 0.022^{+0.004}_{-0.002} \\
VCC0940 & 0.032^{+0.003}_{-0.004} \left(0.025^{+0.002}_{-0.003}\right) & 0.054^{+0.008}_{-0.006} \\
VCC1010 & 0.037^{+0.007}_{-0.008} \left(0.032^{+0.006}_{-0.006}\right) & 0.026^{+0.003}_{-0.003} \\
VCC1695 & 0.081^{+0.011}_{-0.017} \left(0.080^{+0.011}_{-0.017}\right) & 0.041^{+0.004}_{-0.003} \\
VCC1896 & 0.059^{+0.013}_{-0.014} \left(0.059^{+0.013}_{-0.014}\right) & 0.076^{+0.009}_{-0.014} \vspace*{1.5pt}\\
\enddata
\tablecomments{Col. (1): name of the dwarf galaxy. Col. (2): residual light fraction of the bar substructure, measured within the start $(r_{\text{start}})$ and end $(r_{\text{end}})$ radius of the bar. In parenthesis, an alternative measurement starting instead at $r_{\text{start}}=0$. Col. (3): residual light fraction of the spiral arm substructure, measured within the end of the bar $(r_{\text{end}})$ out to two effective radii. If there is no bar detected, the measurement starts instead at $r=0$.}
\end{deluxetable}

We now take advantage of the bar substructure and spiral substructure identification provided by the Fourier analysis, and compute the residual-to-total light fraction in each region, following what was previously done in Section \ref{sec:results}. First, for the bar region, we perform the measurement within the start $(r_{\text{start}})$ and end $(r_{\text{end}})$ radius of the bar. We also provide an alternative measurement starting at $r_{\text{start}}=0$. Second, for the spiral region, we measure within the end of the bar $(r_{\text{end}})$ out to two effective radii. For the galaxies that do not have a bar, no bar measurement is performed, and the spiral region measurement starts instead at $r=0$. The results of the bar and spiral residual light fractions of the dwarf ETG sample are summarized in Table \ref{tab:fourier:bar_sp_rlf}. We find that these two types of substructures contribute a comparable fraction of the total light, with the bar light fraction ranging between $2.8-8.1$\% ($2.2-8.0$\% if measured from $r_{\text{start}}=0$), and the spiral light fraction ranging between $2.2-7.6$\%. The aforementioned bar light fractions are also consistent with the ones obtained through Fourier analysis (see Table \ref{tab:fourier:params}), presenting differences of only a few percent. Additionally, we find no conclusive correlation between the spiral light fraction and the presence of a bar.

\begin{figure}[ht]
\includegraphics[width=\columnwidth]{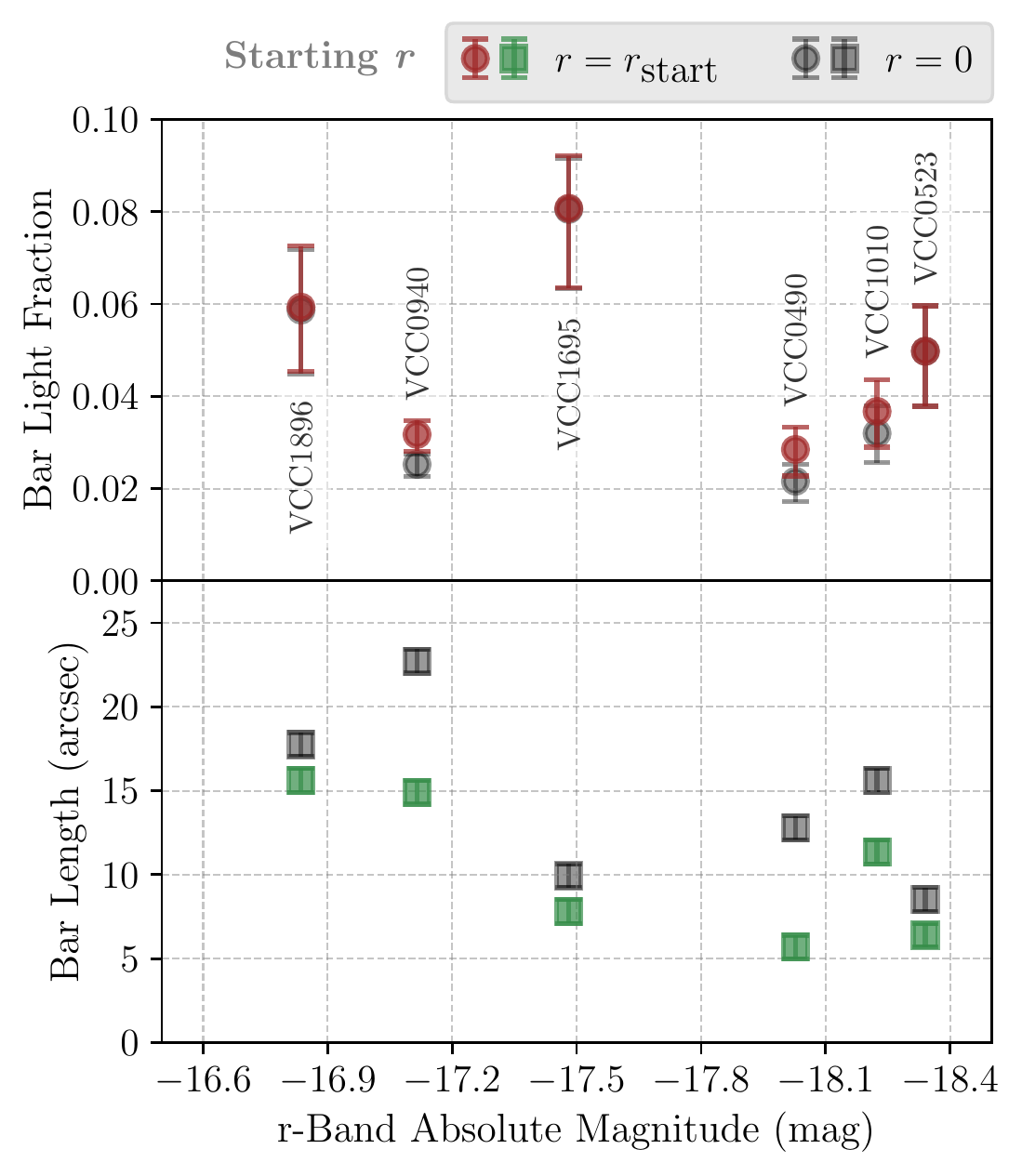}
\caption{Bar light fractions (top panel, circles) and projected bar lengths (bottom panel, squares) of the dwarf ETG sample, as a function of the total absolute magnitude of the galaxy in the $r$-band. The measurements are performed within the bar's start $(r_{\text{start}})$ and end $(r_{\text{end}})$ radius and are shown as colored points. The alternative measurements starting instead at $r_{\text{start}}=0$ are shown as gray points. The error bars of the bar light fraction are given by the 16th and 84th percentiles of the distribution, while the error bars of the bar length are given by the resolution of the Fourier algorithm. Each set of data points is labeled with the name of the dwarf galaxy it corresponds to.\label{fig:fourier:bar_rlf_length}}
\end{figure}
\begin{figure}[ht]
\includegraphics[width=\columnwidth]{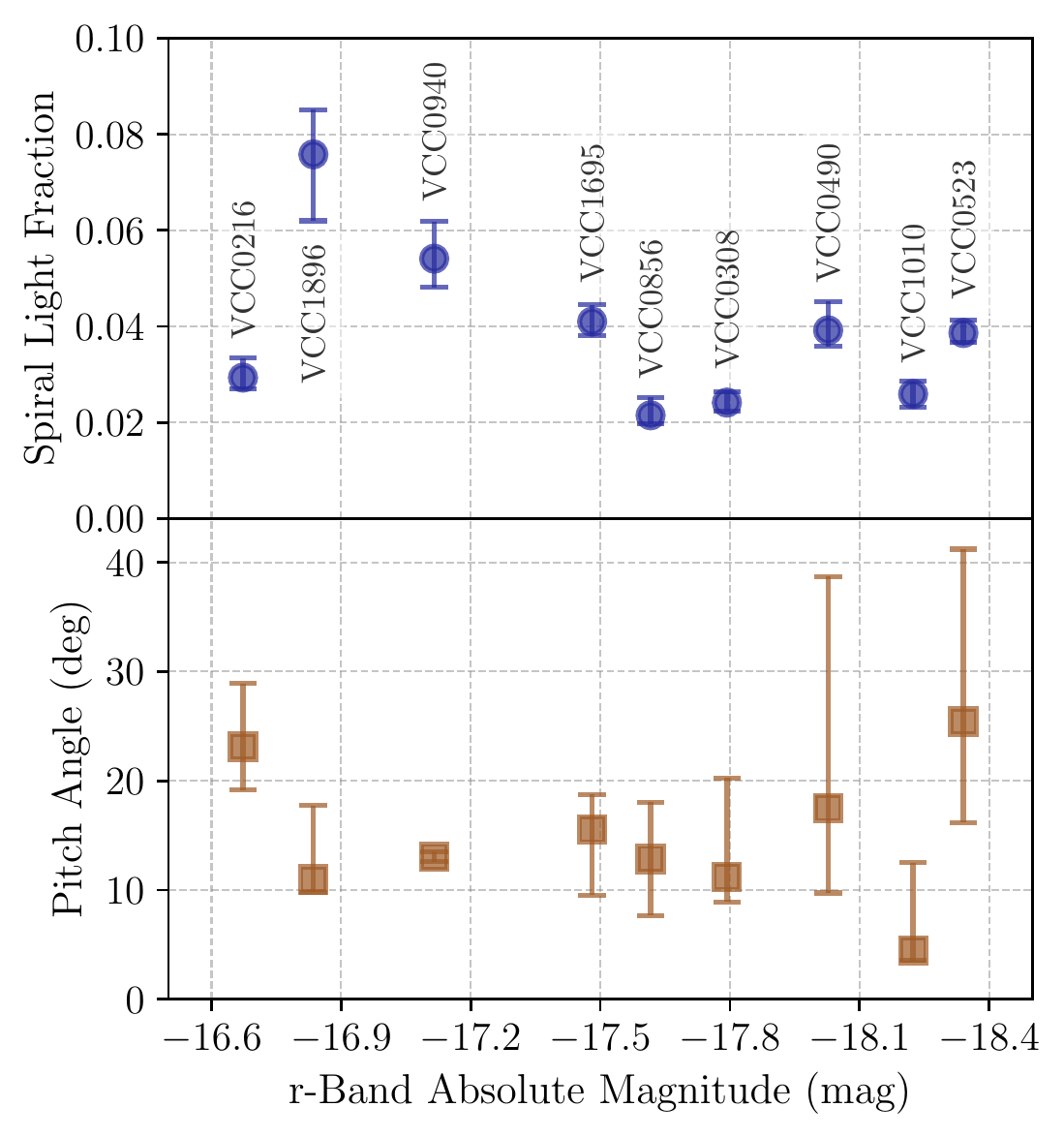}
\caption{Spiral light fractions (top panel, circles) and absolute value of the median projected spiral arm pitch angles (bottom panel, squares) of the dwarf ETG sample, as a function of the total absolute magnitude of the galaxy in the $r$-band. The residual light measurements are performed within the bar's end radius $(r_{\text{end}})$ out to two effective radii. If the galaxy has no identified bar substructure, the measurement starts instead at $r=0$. The error bars are given by the 16th and 84th percentiles of the distributions. Each set of data points is labeled with the name of the dwarf galaxy it corresponds to.\label{fig:fourier:sp_rlf_pitch}}
\end{figure}

As the Fourier analysis has broadened our knowledge about the bar and spiral arm substructures separately, we proceed to search for correlations between the properties of these disk features and the properties of their galaxies. In Figure \ref{fig:fourier:bar_rlf_length} we present the bar properties, the bar residual light fraction and the projected bar length, as a function of the $r$-band absolute magnitude of the galaxies. Similarly, Figure \ref{fig:fourier:sp_rlf_pitch} presents the spiral arm properties, the spiral residual light fraction and the median projected spiral arm pitch angle. It appears that the dwarf ETG sample is relatively homogeneous, as we find no definitive trends between the bar and spiral arm properties and the brightness of the galaxies. \myedit{Nonetheless, there could be a weak indication of decreasing projected bar lengths with increasing galaxy brightness. For the bar length measurements starting at $r_{\text{start}}$, we find a Pearson correlation coefficient $r=0.76$ with a $p$-value $=0.08$.} To ascertain the existence of meaningful correlations at this low-luminosity, low-mass galaxy range, these results would greatly benefit from having a bigger sample of dwarf galaxies to analyze. 

In conclusion, the Fourier analysis has proven to be a useful application of the residual method. \myedit{When applied to the galaxy residual images, it allowed a detailed quantitative analysis of the shape, extension, and light content of their bar and spiral arm substructures.}

\section{Discussion}\label{sec:discussion}

We now address several aspects related to the residual method and our sample of dwarf ETGs. In the following Section \ref{subsec:discussion:strengths}, we describe the strengths of the residual method and how it improves on the weaknesses of the procedure it is originally based on. Then, in Section \ref{subsec:discussion:furtherapps}, we consider the potential of the residual method and how it can be extended to larger and richer data sets. Finally, in Section \ref{subsec:discussion:sims}, we make a link to our companion paper, Brought to Light II: Smith et al. 2021, and \myedit{briefly introduce} how numerical simulations can help to constrain the evolutionary history of dwarf systems that have faint embedded disk substructure.

\subsection{Strengths and Improvements}\label{subsec:discussion:strengths}

The residual method is based on and inspired by the residual image optimization procedure originally presented in Appendix A of \citet{Lisker:2006a}. In comparison, the residual method introduces several improvements, and its main strengths can be summarized into the two following attributes.\\


\noindent \textbf{1. Accuracy.} The residual method is accurate, as it is able to reliably extract both the actual geometry and the light content of the underlying disk substructure of a galaxy image. 

First, we address the appearance of disk features, such as bars and spiral arms. By comparing the unsharp mask images (Figure \ref{fig:data:imgs_gals_uns}) and the residual images obtained with the residual method (Figure \ref{fig:results:imgs_res}), we observe that the geometry of the disk features is preserved. The main difference lies in the innermost galaxy region: the unsharp mask images appear artificially bright \myedit{in the center}, which is an unavoidable, unwanted effect that arises from smoothing \myedit{(as addressed in Section \ref{subsec:rectests:results})}. Instead, the galaxy residual images do not suffer from this central light excess, and their bar substructures adopt an hourglass-like shape. This latter effect has also been observed in other works \citep[e.g.,][]{Barazza:2002,Lisker:2006a}, and is understood as a highly probable indicator of the presence of a bar. Despite this difference, the residual method does a good job in capturing the shape and orientation of any disk features originally revealed through unsharp masking, \myedit{in preserving their pixel-by-pixel resolution,} and in not introducing artificial features. All of this is possible thanks to \myedit{Steps 1 and 2} of the residual method (see Section \ref{subsec:method:steps}), \myedit{in which the galaxy image is smoothed out} and then subjected to a free fitting procedure. First, the smoothing reduces the impact that the disk features have in driving the galaxy geometry. \myedit{Then, the free isophote fitting of the smoothed-out galaxy allows us to capture the overall geometry of its diffuse main body, and construct a galaxy model with it. As a direct consequence, the resulting galaxy residual will contain a geometrically accurate representation of the remainder disk component.} Real galaxies in clusters often present isophotal twists in their light distribution, so being able to properly model the geometrical complexity of their diffuse component is a necessity.

The amount of light contained in the disk features is also accurately extracted. \myedit{As a reminder, this corresponds to the non-smooth, excess light contained in substructures, which the residual method isolates from the rest of the smooth light of the galaxy. While the smooth light is modeled based on the geometry of the isophotes of the smoothed-out galaxy (Step 2), these isophotes are then used to measure the actual brightness of the galaxy itself (Step 3). This way, the actual amount of diffuse light is accurately represented in a model image, consequently leaving the remaining disk substructure light in a residual image.}  In order to test the accuracy of this \myedit{separation process}, we demonstrated in Section \ref{sec:rectests} that the residual method succeeds in recovering the residual light that is manually introduced into mock galaxy images. \myedit{The method also proves to be very sensitive to faint disk substructure}, as we report contributions to the total light that go as low as 2\% for the dwarf ETG sample (see Table \ref{tab:results:res} and Figure \ref{fig:results:plot_res}). \myedit{The robustness of the method} opens the possibility of reliably using it for additional applications, such as the Fourier analysis we performed on the galaxy residual images in Section \ref{sec:fourier}. However, we note that in order to reach this level of accuracy very deep images were required, as the accuracy of any measurement depends on their uncertainties, and hence on the S/N of the data. Thus, such an analysis is possible provided that data of similar depths \textendash comparable to the depth of modern surveys, like the NGVS \citep{Ferrarese:2012}\textendash\ are available.

In comparison, the residual image optimization procedure of \citet{Lisker:2006a} adopts a \myedit{fixed instead of a free fit approach, by imposing a single value for the ellipticity and position angle during the isophotal fitting of the galaxy image. In Section \ref{subsec:rectests:results}, we showed that a fixed fit approach fails to properly describe the light distribution of a galaxy when it is geometrically more complex. Furthermore, their approach does not include a smoothing procedure, which is a new additional step in our own residual method. All of these improvements we have incorporated} are what ensures that both the geometry and the brightness of the galaxy components are captured faithfully. It should also be taken into consideration that the study of \citet{Lisker:2006a} makes use of much shallower SDSS images, meaning that the accuracy of their measurements has intrinsically higher uncertainties when compared to ours.\\

\noindent \textbf{2. Adaptability.} The residual method is adaptable, as both the iterative aspect of the method and the configuration of its free parameters can be tuned on a case-by-case basis.

First, we refer to the iterative procedure, corresponding to Step 7 of the residual method 
(described in Section \ref{subsec:method:steps} and illustrated in Figure \ref{fig:method:schematic_its}). \myedit{The number of iterations that are required depend specifically on the characteristics of the particular galaxy being examined. This is possible thanks to our introduction of a stopping criterion, which dynamically adjusts the number of iterations based on the case at hand. This case-by-case adaptation is what ultimately allows an accurate determination of the amount of light contained in disk features.} As the iterative procedure gradually separates the diffuse and disk components into a model and a residual image, respectively, it is of utmost importance that the iterations stop once the decomposition is complete. If the iterations were to continue \myedit{arbitrarily longer, additional light would begin to be shifted from the diffuse to the disk component, resulting in artificially brighter disk features and consequently in an inaccurate estimation of their light content. For this reason, we iterate only as long as the stopping criterion deems it necessary.}

Similarly, the parameter configuration of the residual method is also dependent on the characteristics of the data. There are two free parameters that must be set: the smoothing kernel size and the sampling step size. As described in Section \ref{subsec:method:config}, factors such as the S/N, the seeing and PSF, the resolution, and the average width and relative strength of the disk features are to be taken into account. To incorporate variations throughout the data set, the parameter configuration adopts a range of possible smoothing and sampling sizes, specifically tuned to the data properties. Then, this whole range is applied to all the galaxy images. This way, when quantifying the disk substructure light, any case-by-case variations are translated into an uncertainty range, allowing a consistent measurement throughout the sample.

In comparison, even though the residual image optimization procedure of \citet{Lisker:2006a} introduces the concept of iterations, the number of iterations they adopt is arbitrary. They do not implement a stopping criterion, \myedit{and instead they iterate all the galaxies in their sample a fixed total of ten times. The residual method improves on this: by establishing a clear stopping criterion}, we are required to iterate only between two to five times, which is less than half the number of iterations used in \citet{Lisker:2006a}. \myedit{For our galaxy images, iterating ten times would have caused the residuals to suffer from significant under-subtraction, which would make them artificially brighter. For this reason, we believe that the measurements of \citet{Lisker:2006a} could be} systematically biased towards higher values. For the three galaxies that we have in common (VCC0308, VCC0490, and VCC0856), they report residual light fractions that are on average three times brighter than the ones measured in this work (see their Table 2 and our Table \ref{tab:results:res}). However, we also have to take into account that their measurements are based on images that have a different depth, seeing, and S/N compared to ours, so there may be other potential reasons contributing to this discrepancy. Nonetheless, thanks to the flexibility of our iterative procedure, we believe that \myedit{we carry out a robust quantification of the actual amount of light contained in the disk substructure of the dwarf ETG sample.}

\subsection{Further Applications}\label{subsec:discussion:furtherapps}

As shown in this work, the main objective of the residual method is to reliably separate the diffuse and disk components of a galaxy image. Once the disk substructure is extracted and isolated in a galaxy residual image, it can then be used to estimate its relative contribution to the total galaxy light. However, this also opens the possibility of subjecting these residual images to a variety of additional analyses. As an example, \myedit{in Section \ref{sec:fourier}, we showed what can be learned about the faint embedded disk substructure} when a Fourier analysis of the residual images is performed. This allowed us to identify and characterize in detail the bar and spiral arm features of the galaxies. The residual method thus holds the potential of having further applications, \myedit{which could help advance our understanding of the nature of the disk substructure present in dwarf ETGs like the ones in our sample}.

The dwarf ETG sample that was analyzed, however, presented two main limitations: low number statistics and a lack in the amount of information that we could derive from it. On the one hand, the data set is a small one, consisting of only nine dwarf galaxies. As shown both through the residual method results in Section \ref{sec:results} and the Fourier analysis results in Section \ref{sec:fourier}, the small sample size hinders our ability of \myedit{drawing meaningful correlations between derived quantities and global galaxy properties}. Thus, the next logical step would be to apply the residual method to a bigger sample of dwarf galaxies. On the other hand, our data set also provides a limited amount of information, as it only consists in imaging data that are restricted to one band. With deep multi-band imaging, it would be possible to carry out integrated color and color profile analyses of the dwarf galaxies \citep{Urich:2017}. Combined with stellar population synthesis models \citep[e.g., MILES;] []{Falcon-Barroso:2011}, the age and metallicity information of the galaxies could then be estimated. \myedit{By applying the residual method to such a data set, it would be possible to isolate the disk substructure and quantify its relative contribution to the total galaxy light in each band separately. Going one step further, we could then derive the integrated colors, color profiles, and stellar population information of the diffuse and disk components.} By comparing the ages and metallicities of the two components, we \myedit{would be taking a step forward in trying to constrain the formation and evolution of such dwarf systems with embedded disk substructure. Deep multi-band imaging can be provided by modern surveys of nearby galaxy clusters, such as NGVS \citep{Ferrarese:2012} in Virgo, or NGFS \citep{Munoz:2015} and FDS \citep{Iodice:2016} in Fornax, and also by future large-scale sky surveys, such as LSST \citep{Ivezic:2019}.}

Last but not least, the residual method is not limited to the particular application we have given it in this work. In theory, it should be possible to apply it to any galaxy image that can be separated into two components; \myedit{a bright, smooth, axisymmetric component, and a faint, non-smooth, non-axisymmetric component}. Thus, from the high-mass end of giant elliptical galaxies with dust lanes to the low-mass end of dwarf spheroidal galaxies with underlying substructure or even dwarf irregular galaxies, the residual method should constitute a valid approach for decomposing and analyzing their structural features. However, the inherent limitations of the method should always be kept in mind, as it is designed to work properly in the case that faint substructure is embedded in a much brighter diffuse body. This means that it is not originally intended to be used, for example, for quantifying the substructure light in giant spiral galaxies, where the spiral arm light is dominant.

\subsection{Link to Simulations}\label{subsec:discussion:sims}

In our companion paper, Brought to Light II: Smith et al. 2021, we present high resolution numerical simulations of dwarf systems that are being subjected to tidal harassment by a galaxy cluster potential. On the one hand, the galaxy cluster corresponds to a Virgo-like, time-evolving analytical model, in which the tidal fields of the main cluster halo and of the individual galaxy halos have been extracted from a cosmological simulation and replaced by analytical potentials. On the other hand, the model galaxies being thrown into the cluster potential consist of particles that inhabit a Navarro-Frenk-White \citep[NFW;][]{Navarro:1996} dark matter halo and a stellar exponential disk.

This companion study focuses on the conditions under which disk features are triggered in the stellar disk of the model galaxies. We conclude that two main factors influence the formation and evolution of disk features: \textit{(a)} how plunging the orbit is in which the model galaxy falls into the cluster, and \textit{(b)} the level of rotational support of the stellar disk. As the main result, we find that disk features form in galaxies that experience close pericenter passages to the cluster core and that have a component of their stellar disk that is highly rotationally supported. Notably, the wide range of disk features exhibited by these simulated galaxies closely resemble the appearance of the disk features we detect in our observational dwarf ETG sample. Thus, we formulate the hypothesis that the disk substructure present in our dwarf ETGs could have been tidally triggered by the Virgo cluster potential.

While it has been found that dwarf ETGs can indeed be rotationally supported \citep{Toloba:2009}, they usually have low to medium rotational support, \myedit{which contradicts} the high rotational support required by our simulations. Therefore, for the tidal triggering of disk features to be possible, a fraction of the light of the galaxy is required to be in the form of a thin, dynamically cold stellar disk, while the remaining light can be in the form of a thick, dynamically hot stellar disk. To test this conjecture, in our companion paper we construct a mixed galaxy model, in which 20\% of the stellar mass is in a thin disk component and 80\% is in a thick disk component. We find that while the thick disk component remains featureless, the thin disk component \myedit{is still sensitive to the cluster potential and reacts to it by forming disk features.} Consequently, it is plausible that our observed dwarf galaxies could have a faint, thin disk component embedded in a much brighter, thick disk component. If the two components were to share similar properties, such as their overall color, then they would be mostly indistinguishable under normal, non-harassment circumstances, as both would be smooth in appearance. Therefore, we propose that the thin disk component could reveal itself through the formation of disk features, such as bars and spiral arms, that are triggered when the galaxy is subjected to harassment along a plunging orbit. In this scenario, the presence of faint disk substructure in our dwarf ETG sample could be an indication that a small fraction of their light is contained in a thin and highly rotationally supported disk, which may have been tidally triggered by the Virgo cluster potential.

For a full interpretation of the observational and simulated results, and a detailed assessment of the possible origins of the disk substructure in cluster dwarf ETGs, please refer to our companion paper, Brought to Light II: Smith et al. 2021.


\section{Summary}\label{sec:summary}

In this work we introduce a newly developed procedure, \myedit{the ``residual method'', which aims to identify and extract the disk features, such as bars and spiral arms, from a galaxy image}. The development and testing stages of the method were carried out based on a deep imaging sample of dwarf ETGs from the Virgo cluster, \myedit{characterized by the presence of faint} underlying disk substructure that lies hidden in the much brighter diffuse main body of the galaxies. The residual method aims at gradually separating the dwarf galaxy images into two distinct components: a bright diffuse component, which is placed in a galaxy model image, and a much fainter disk component, which is isolated in a galaxy residual image. \myedit{We describe in detail the constituting steps and the parameter configuration of the method, such that it can be easily reproduced and adapted for its application to different data sets.}

By quantifying the relative contribution of the disk component to the total galaxy light in our dwarf ETG sample, we find that the disk features are indeed very faint, as they constitute between $2.2-6.4$\% of the galaxy light within two effective radii. In order to assess the reliability of these results, we subject the residual method to diagnostic tests, and evaluate the accuracy with which it is able to recover different amounts of disk substructure light that we have manually introduced in mock galaxy images. We prove that the method is well behaved, and that it is much better suited for the purpose of quantifying faint disk substructure than the simpler alternative approaches we compare it with.

In order to showcase a potential application of the residual method, we perform a Fourier analysis of the dwarf ETG sample. \myedit{We find that by applying a Fourier decomposition directly to the galaxy residual images, it is possible to clearly identify and characterize the disk features that normally lie hidden in the galaxies}. Thus, we are able to separate their bar and spiral arm regions, measure their individual contributions to the total galaxy light, and derive characteristic quantities such as the bar lengths and the spiral arm pitch angles.

Finally, we discuss the main strengths of the residual method. \myedit{These consist in} the accuracy with which it extracts the light contained in the disk features of a galaxy, and in the way it adapts to the properties of the data on a case-by-case basis. However, we point out that the application of the method would greatly benefit from having a bigger galaxy sample with \myedit{deep} multi-band imaging, which would allow us to better interpret the results and test correlations between derived quantities and global galaxy properties. We also make a link to our companion paper, Brought to Light II: Smith et al. 2021, and address how numerical simulations can help us to constrain the \myedit{formation and evolution} of disk substructure in cluster dwarf ETGs. \myedit{The studied scenario proposes that} these dwarf systems could contain a faint, thin, and highly rotationally supported stellar disk, in which disk features are triggered through the tidal interaction with the Virgo cluster potential.

\acknowledgments
\myedit{We thank the anonymous referee for a thoughtful report that helped improve the manuscript.} J.M. acknowledges the support of the Deutscher Akademischer Austauschdienst (DAAD) through a doctoral scholarship, and the funding provided by Universit\"at Heidelberg and the Max-Planck-Institut f\"ur Astronomie. We also acknowledge support from the European Union’s Horizon 2020 research and innovation program under the Marie Sklodowska-Curie grant agreement no. 721463 to the SUNDIAL ITN network.

\facility{Max Planck:2.2m (WFI).}

\software{Astropy \citep{Astropy:2013, Astropy:2018}, IRAF \citep{IRAF:1986, IRAF:1993}, Lucidchart (\href{https://www.lucidchart.com/}{www.lucidchart.com}), NumPy \citep{Numpy:2006, Numpy:2011}, Matplotlib \citep{Matplotlib:2007, Matplotlib:2020}, pandas \citep{pandas:2010, pandas:2020}, SAOImage DS9 \citep{ds9:2003, ds9:2019}, scikit-image \citep{skimage:2014}, SciPy \citep{Scipy:2020}, STSDAS \citep{stsdas:1994}.}


\bibliographystyle{aasjournal}{} 
\bibliography{residual_method}

\end{document}